\documentclass[a4paper,fleqn,usenatbib]{mnras}
\usepackage[T1]{fontenc}
\usepackage{ae,aecompl}

\usepackage{amsmath}
\usepackage{graphicx}	
\usepackage{lscape}
\usepackage{enumitem}
\usepackage{indentfirst}

\usepackage{amssymb}
\usepackage[dvipsnames]{xcolor}
\usepackage[flushleft]{threeparttable}

\newcommand {\bc}{\begin {center}}
\newcommand {\ec}{\end {center}}
\newcommand {\be}{\begin {equation}}
\newcommand {\ee}{\end {equation}}
\newcommand {\beq}{\begin {eqnarray}}
\newcommand {\eeq}{\end {eqnarray}}

\newcommand {\ergs}{{\rm erg\ \rm s^{-1}}}
\newcommand {\comment}[1]{}

\def\lbar {\lambda\hskip-5pt\raise3pt\hbox {--}}
\def\lbr {\lambda\raise2pt\hbox {\hskip-4pt{\scriptsize --}}_\C}

\renewcommand{\d}{{\rm d}}

\renewcommand{\d}{{\rm d}}

\title[Luminosity and pulsed fraction in bright XRPs]
{
Apparent luminosity and pulsed fraction affected by gravitational lensing of accretion columns in bright X-ray pulsars
}
\author[I. D.~Markozov and A. A.~Mushtukov] 
{Ivan~D.~Markozov$^{1}$\thanks{E-mail: markozoviv@mail.ru (IDM)}
and
Alexander~A.~Mushtukov$^{2}$\thanks{E-mail: alexander.mushtukov@physics.ox.ac.uk (AAM)}
\\ 
$^1$ Ioffe Institute, Politekhnicheskaya 26, St Petersburg 194021, Russia \\ 
$^2$ Astrophysics, Department of Physics, University of Oxford, Denys Wilkinson Building, Keble Road, Oxford OX1 3RH, UK\\
} 

\pubyear{2023}

\begin{document}
\label{firstpage}
\pagerange{\pageref{firstpage}--\pageref{lastpage}}
\maketitle

\begin{abstract}

The luminosity of X-ray pulsars is their key parameter determining the geometry and physical conditions of the accretion flow both on the spatial scales of a binary system and on much smaller scales of emitting regions located close to the stellar surface.
Traditionally, the luminosity of X-ray pulsars is estimated out of the X-ray energy flux averaged over the pulsed period and the estimated distance to the source. 
Due to the anisotropy of X-ray emission, the luminosity estimated on the base of the observed pulse profile can differ from the actual one.
Super-critical X-ray pulsars with accretion columns are of particular interest because the X-ray flux from columns is a matter of strong gravitational lensing by a neutron star.
Using toy model of an accretion column, we simulate beam patterns in super-critical X-ray pulsars, construct theoretical pulse profiles for different geometries and mutual orientations of pulsars and distant observers and show that despite strong light bending, the typical deviation of the apparent luminosity from the actual one is $\sim 20\%$ only, 
and in 90\% of cases, the apparent luminosity
$0.8 L\lesssim L_{\rm app}\lesssim 1.25 L$.
However, the shape of the pulse profiles is strongly affected by the geometry of the emitting region. 
We show that the appearance and growth of accretion columns tend to be accompanied by an increase of observed pulsed fraction, which is in agreement with the recent observations of bright X-ray transients. 
\end{abstract}


\section{Introduction}

X-ray pulsars (XRPs) are close binary systems where one of the components is a strongly magnetised neutron star (NS) accreting matter from the companion star (see \citealt{2022arXiv220414185M} for the recent review).
Typical magnetic field strength at the surface of a NS in XRPs is $\sim 10^{11}-10^{13}\,{\rm G}$, which is confirmed by detection cyclotron lines in spectra of many XRPs (see, e.g., \citealt{2019A&A...622A..61S}). 
Strong magnetic field affects geometry of an accretion process directing accretion flow towards small regions located close to magnetic poles of a NS (typical area is $\sim10^{9}\,{\rm cm}$). 
In the vicinity of NS magnetic poles accreting matter loses its kinetic energy, which results in emission of photons predominantly in X-ray energy band (see, e.g., Chapter 6 in \citealt{2002apa..book.....F}). 
Misalignment between rotational and magnetic axes of a NS leads to appearance of pulsation and phenomenon of XRP. 
Apparent luminosity of XRPs covers almost ten orders of magnitude from $\sim 10^{32}\,\ergs$ up to $\sim 10^{41}\,\ergs$.
The brightest XRPs belong to the class of pulsating ULXs (see \citealt{2021AstBu..76....6F,2023NewAR..9601672K}) or bright Be X-ray transients \citep{2011Ap&SS.332....1R}.

The estimation of accretion luminosity in XRPs is based on the measurement of the X-ray energy flux averaged over the pulse period $P$ of XRP
\beq \label{eq:F_ave}
F_{\rm ave}=\frac{1}{P}\int\limits_0^PF(t){\rm d}t.
\eeq 
an estimated distance to the source $D$, and assumptions about energy distribution in the spectra:
\beq\label{eq:L_app}
L_{\rm app}=4\pi D^2 F_{\rm ave}.
\eeq 
For many XRPs located in our galaxy, distances remain the major cause of uncertainty, being often known with errors as large as 50 per cents \citep{2021AJ....161..147B}.
The pulse profile and the average flux measured by a distant observer are determined by the beam pattern and geometry of NS rotation. 
The beam patterns is affected by the geometry of the emitting region and expected to be dependent on the mass accretion rate onto a NS surface \citep{1973A&A....25..233G}. 
The geometrical parameters of NS rotation in the observer's reference frame (i.e., inclination of rotating NS and magnetic obliquity, i.e., the angle between the spin axis and magnetic axis) are purely known.
However, there is some progress in determination XRP geometry on the base of data obtained with X-ray polarimeters (see, e.g., \citealt{2022NatAs...6.1433D,2023A&A...677A..57D,2022ApJ...941L..14T,2023arXiv230206680T,2023arXiv230317325M,2023arXiv230515309S,2023A&A...675A..29M}).

Depending on the mass accretion rate one would expect different geometry of the emitting region at the magnetic poles. 
At relatively low mass accretion rates ($\dot{M}<10^{17}\,{\rm g\,s^{-1}}$) and, therefore, accretion luminosity, radiative force does not affect much the dynamics of accretion flow above NS surface and accretion process results in hot spots located close the magnetic poles of a star \citep{1969SvA....13..175Z}. 
If accretion luminosity reaches the critical value $L_{\rm crit}\sim 10^{37}\,\ergs$ radiation pressure becomes high enough to stop incident flow in a radiation dominated shock \citep{1976MNRAS.175..395B,2015MNRAS.447.1847M}. 
The higher the mass accretion rate and luminosity, the higher the shock above the surface of a NS. 
Below the shock region, matter slowly settles onto the stellar surface forming a so-called accretion column (see e.g., \citealt{1975PASJ...27..311I,1976MNRAS.175..395B,1981A&A....93..255W,2015MNRAS.454.2539M,2022MNRAS.515.4371Z}). 
Height of the column can be comparable to the size of a NS and, therefore, its geometry strongly affects the beam pattern of X-ray emission \citep{1988ApJ...325..207R,2001ApJ...563..289K,2018MNRAS.474.5425M,2020PASJ...72...34I}.
X-ray emission from the side walls of the accretion column is expected to be largely beamed towards NS surface due to relativistic effects \citep{1976SvA....20..436K,1988SvAL...14..390L}.
A fraction of X-ray emission is intercepted by the NS surface and reprocessed by its atmosphere \citep{2013ApJ...777..115P,2015MNRAS.452.1601P,2019Ap.....62..129G,2021A&A...655A..39K}. 
Reprocessed emission forms an additional part of the emission pattern. 

At extremely high mass accretion rates ($>few\times 10^{19}\,{\rm g\,s^{-1}}$), accretion disc in XRPs loses material due to the radiation driven outflows \citep{1999AstL...25..508L,2007MNRAS.377.1187P}.
Outflows can collimate X-ray radiation from the central NS affecting the apparent luminosity \citep{2009MNRAS.393L..41K,2017MNRAS.468L..59K} and influence pulsations \citep{2023MNRAS.518.5457M}.
These type of sources is beyond the scope of the paper.

In this paper, we focus on super-critical XRPs, where the mass accretion rate is large enough to cause the formation of accretion columns above the surface but is still low to launch the outflows from the disc, i.e. $10^{17}\,{\rm g\,s^{-1}}\lesssim \dot{M}\lesssim few\times 10^{19}\,{\rm g\,s^{-1}}$.
The complicated geometry of emitting regions at high mass accretion rates and gravitational lensing of radiation from the accretion column affect pulse profile formation.
Under these conditions, one could expect significant deviations of the apparent luminosity estimated on the base of the observed pulse profiles from the actual luminosity of XRPs.
In this paper, we reproduce statistical distributions of bright XRPs over the apparent luminosity and estimate the typical difference between actual and apparent luminosity.
We focus particularly on the case of super-critical XRPs with accretion columns.
We account for the effect of gravitational light bending \citep{1983ApJ...274..846P,2002ApJ...566L..85B,2020A&A...640A..24P} and emission pattern composed of the direct emission from hot spots in the case of sub-critical accretion and accretion columns and emission reprocessed by the atmosphere of a NS in the case of super-critical mass accretion rates.
The pulse profiles and the apparent luminosity depend on the observer's viewing angle to the XRP.
Because the observer's viewing angle is not usually known, we apply statistical analysis to estimate possible deviations between the apparent and actual luminosity.

\section{Model set up}
\label{sec:PhysCond}

\begin{figure*}
	\centering 
	\includegraphics[width=14cm]{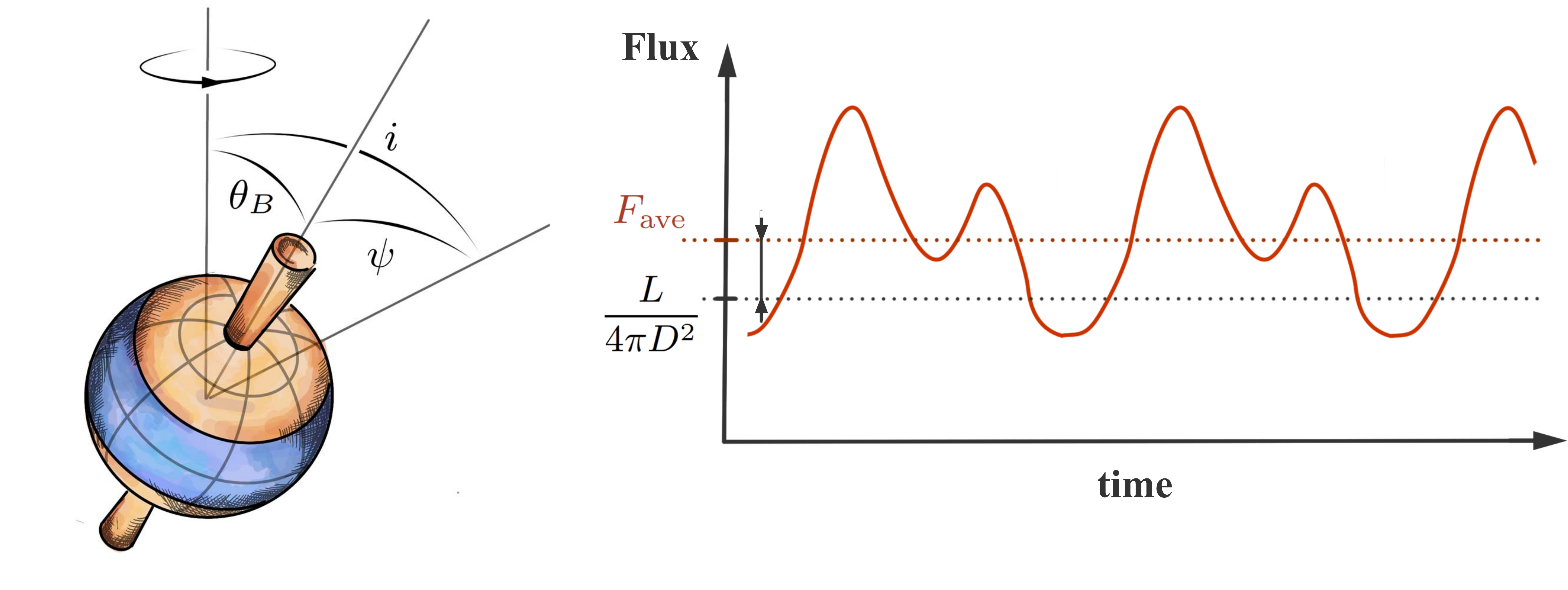} 	
	\caption{
    Schematic illustration of emission region geometry at super-critical mass accretion rates in XRPs.
    The flux detected by a distant observer is composed of the direct flux from accretion column and the flux intercepted and reprocessed by stellar surface.
    Rotation geometry of a NS in observer's reference frame is determined by two angles: NS inclination $i$ and the magnetic obliquity $\theta_B$.
        The flux averaged over the pulsation period $F_{\rm ave}$ is dependent on the viewing angle, beam pattern and parameters of NS rotation. 
    It also can be differ from the isotropic flux of a NS expected at its luminosity $L/(4\pi D^2)$.
}
	\label{pic:scheme} 
\end{figure*}

We consider a spherically-symmetric NS of radius $R$ and mass $M$.
The magnetic field of the NS is assumed to be dominated by the dipole component. 
In this case the material from the accretion disc or stellar wind reaches NS surface in small regions located close to the magnetic poles of the star. 
We take the geometrical size of an accretion channel base $r_c$ as a parameter. 
It depends on the magnetic field strength $B$, mass accretion rate $\dot{M}$ and geometry of accretion flow, but this dependence is weak: $r_c\propto \dot{M}^{1/7}B^{-2/7}$ \citep{2015MNRAS.447.1847M}.
Accretion column height $H$ in a super-critical regime of accretion is governed by the mass accretion rate \citep{1976MNRAS.175..395B} and also used as a parameter in our simulations too.

Photons emitted by hot spots or accretion columns propagate in a space curved by a NS. 
We limit our analysis to the case of Schwarzschild metric, which is a good approximation for the case of XRPs since they have relatively long spin periods (typically, $P>0.1\,{\rm s}$).
The fraction of X-ray radiation emitted by accretion column is intercepted and reprocessed by the atmosphere of a NS \citep{2013ApJ...777..115P,2018MNRAS.474.5425M,2021A&A...655A..39K}. 
We do not consider spectral transformation of X-ray radiation reprocessed by the atmosphere and assume that the flux emitted by the unit area at the NS surface is equal to the total flux reaching the surface at a given latitude. 
A distant observer detects both direct emission from accretion columns and emission reprocessed by the stellar atmosphere, and both components form the pulse profile. 

The pulse profiles constructed under the discussed assumptions are used to get the apparent luminosity (\ref{eq:L_app}) of XRP, which can be further compared to the actual one used in the model. 
The apparent pulse profile and luminosity of XRP depend on the mutual orientation of a rotating NS and a distant observer.
Rotation of the NS in XRP is determined by the magnetic obliquity $\theta_B$, i.e. the angle between the magnetic and rotation axes of a NS.
The observer's position is specified by the inclination, i.e., the angle
$i$  between the rotation axis and the line of sight.
Investigating statistical features of apparent luminosity we assume the distant observers are randomly oriented in the reference frame of a binary system. 
The angles $i$ and $\theta_B$ take random values from the interval $[0,\pi]$ with probability density $\rho(\theta)\propto\sin{\theta}.$

\subsection{Sub-critical case}

Under the condition of sub-critical mass accretion rates ($\dot{M}\lesssim 10^{17}\,{\rm g\,s^{-1}}$), emission is produced by hot spots located at the opposite magnetic poles of a NS. 
{The geometry of hot spots is dependent on the global geometry of accretion flow. 
In the case of accretion from the stellar wind, the spots have round shape, while accretion from the disc results in a ring shape of the spots.
Under the conditions of strong magnetic fields ($B\gtrsim 10^{12}\,{\rm G}$) typical for XRPs, the geometrical size of hot spots is small $\lesssim 10^5\,{\rm cm}$ and their geometry does not affect significantly the process of pulse formation.
Thus, we treat hot polar spots like a segments of sphere with an opening angle} 
\beq 
\theta_{\rm pol}\approx\left(\frac{R}{R_{\rm m}}\right)^{1/2}\sim 0.1\text{~rad},
\eeq 
where $R_{\rm m}$ is the magnetospheric radius.
The specific intensity of X-ray radiation leaving the NS surface is considered to be independent on the coordinate within the spot, and
the angular distribution of specific intensity at the stellar surface is modelled as
\beq
\label{eq:polar_cap_distr_fr}
	 I(\alpha_{\rm s})\propto 1+b\cos\alpha_{\rm s}
\eeq
where $\alpha_{\rm s}$ is the angle between the normal to the surface and photons momentum, and $b>-1$ is the parameter of the model describing the geometry of a beam pattern.
Parameter $b<0$ corresponds to the case of beam pattern suppressed along the normal to the NS surface, like it is expected at extremely low levels of accretion (see, e.g., \citealt{2021MNRAS.503.5193M,2021A&A...651A..12S}). 
The value $b=2$ corresponds to the case of optically thick atmosphere with dominating Thomson scattering (\citealt{1963trt..book.....S}).

\subsection{Super-critical case}

In the case of super-critical accretion, the emitting regions are given by two accretion columns of height $H$ above the magnetic poles of a star. 
{Accretion column height can be comparable to the radius of a NS \citep{1976MNRAS.175..395B,2022arXiv220712312A}. 
Column height is related to the accretion luminosity of a NS: the larger the luminosity, the higher the accretion column. 
The relation between the column height and luminosity can be approximated as (see, e.g., \citealt{2015MNRAS.454.2539M})
\beq\label{eq:L2H_relation}
\frac{L}{L_{\rm Edd}}\approx 40\,\left(\frac{\kappa_{\rm T}}{\kappa_\perp}\right)q\left(\frac{H}{R}\right), 
\eeq
where $\kappa_{\rm T}\approx 0.34\,{\rm cm\,g^{-1}}$ is the Thomson electron scattering opacity, $\kappa_\perp$ is the effective Compton scattering opacity in a strong magnetic field in the direction orthogonal to the field lines, and 
\beq\label{eq:f}
q(x)\equiv \log(1+x)-\frac{x}{1+x}.
\eeq
According to some models, accretion columns tent to be unstable at $H>R$ \citep{2015MNRAS.454.2539M}.
The columns in our simulations are treated as cylinders of a given radius $r_c\ll R$.
This approximation does not account for upward expansion of accretion column due to the dipole geometry of NS magnetic field.
However, this approximation is reasonable under conditions of small base of accretion column in XRPs and limited column height (we consider $H<R$).}
To describe the distribution of accretion column brightness over its height and the angular distribution of X-ray radiation leaving the walls of the accretion column we introduce functions  $g(h)$ and $f(\alpha,\phi)$ respectively. 
Here $h$ is the height above NS surface, $\alpha$ is the angle between the direction of the photon momentum and the column surface, and $\phi$ is the azimuth angle. 

The distribution of accretion column luminosity over its height $g(h)$ was considered in three cases:
(a) uniform luminosity distribution $g(h)=1/H$;
(b) the distribution implying brightening of accretion columns towards their base  (see, e.g., \citealt{2015MNRAS.454.2539M}):
\beq
\label{height_distr_fr}
	 g(h)\propto\frac{H-h}{R+h};
\eeq 
(c) the distribution $g(h)\propto h$, which corresponds to a possible luminosity growth to the column top.

The radiation energy from unit surface area located at the height $h$ into the unit solid angle of the direction $(\alpha_0,\phi)$ is given by $F_\mathrm{out}f(\alpha_0,\phi)g(h)$, where $F_\mathrm{out}$ is the 
the flux leaving the accretion column wall.
The angular distribution of X-ray flux leaving the walls of the accretion column can be affected by high velocity of flow at the edges of the accretion channel due to relativistic effects \citep{1976SvA....20..436K,1988SvAL...14..390L}.
As a result, radiation can be strongly beamed toward the NS surface.
Following \citealt{2013ApJ...777..115P} we describe the angular distribution of X-ray flux at the walls of accretion columns as
\beq
\label{angle_distr_column}
f(\alpha_0,\phi)\propto\frac{3D^4}{7\pi\gamma}
\left(1+2D\sin\alpha_0\cos\phi\right)
\sin^2\alpha_0\cos\phi
\eeq
where $D=[\gamma(1-\beta\cos{\alpha_0})]^{-1}$ is the Doppler factor, $\gamma=(1-\beta^2)^{-1/2}$ is the Lorentz factor and $\beta=v/c$ is the dimensionless velocity in units of the speed of light. 
The velocity at the edges of the accretion cavity is a model parameter. 
Within the framework of this work two limiting cases are considered: 
(i) zero velocity of the walls  $\beta=0$, and 
(ii) free-fall velocity $\beta=\beta_{\rm ff}=[R_{\rm Sh}/(R+h)]^{1/2}$, where $R_{\rm Sh}=2GM/c^2\simeq 2.95\times 10^5 \,{\rm cm}$ is the Schwarzschild radius.	

A fraction of X-ray photons produced by the accretion column is intercepted by the NS surface and reprocessed.
We assume that the energy re-emitted locally by the NS surface is equal to the incident X-ray energy and the specific intensity of reflected radiation doesn't depend on the direction.

Propagation of photons both from the accretion column and from the illuminated surface of the NS is a subject of gravitation bending.

\section{Numerical approach}
\label{sec:alg}

The work is divided into the following stages:
\begin{enumerate}[leftmargin=12pt]
\item 
Getting the angular distribution of the flux.\\
For a given actual accretion luminosity and geometry of the emitting regions, we get the angular distribution of X-ray flux from a NS in the reference frame related to the dipole component of stellar magnetic field. 
Because of the cylindrical symmetry, the flux dependent only on the angle $\psi$ between the line of sight and magnetic axis of a NS.
The angular distribution of the flux is obtained using the ray tracing method. 
{The sphere of infinitely distant observers around the NS broke up into rings of equal angular thickness ${\rm d}\psi$, measured from one of the NS poles. The NS itself is also split into rings in latitude, measured from the same magnetic pole. Due to the symmetry relative to the equator plane of the NS, the choice of a specific pole as a reference point does not affect the results.}\\
In the super-critical case, the accretion column is uniformly split into $N_{\rm col}$ pieces over height (in the presented simulation we use $N_{\rm col}=1000$).  
Then for each piece the uniform set of angles $\alpha_0$ between the normal to the column surface and the direction of the beaming ray and azimuth angles $\phi$ is constructed. 
Rays are launched into these angles. 
Each of them carries an energy $E\propto g(h)f(\alpha_0,\phi)$. 
The final direction of the ray propagation is calculated accounting for the gravitational light bending (see \ref{App:GR}). 
If the ray goes to infinity, its energy corrected for gravitational redshift is added to the corresponded latitude $\psi$ on the sphere of infinity distant observers. 
Similarly, if the ray intersects the stellar surface, the area of the surface where the ray landed gets the energy of the ray corrected for the gravitational redshift. 
We perform these procedures for both accretion columns. 
After that, the ray tracing of the radiation reflected by the NS surface is conducted. 
We assume that the energy re-emitted locally by the NS surface is equal to the incident X-ray energy and the specific intensity of reflected radiation does not depend on the direction.
\item 
Getting the pulse profiles.\\
For a given displacement of the observer in respect to the XRP, we construct the expected apparent pulse profile.
To get the pulse profile we calculate variations of the observer's viewing angle $\psi$ during the pulse period.
The angle $\psi$ is dependent on the orientation of rotational axis in respect to the distant observer given by the inclination of a NS $i$, magnetic obliquity $\theta_B$ and pulsation phase {$\varphi_{\rm p}\in[0;2\pi]$} (see Fig.\,\ref{pic:scheme}), and can be calculated as
\beq
\cos{\psi}=\cos i\cos\theta_B +\sin i\sin\theta_B\cos\varphi_{\rm p}
\eeq
Hence at fixed angles $\theta_B$ and $i$ the pulse profile can be obtained based on the flux distribution over angles $\psi$, which has been obtained at the previous step.
\item 
Getting the apparent luminosity.\\
On the base of the obtained pulse profiles, we get an estimation of the average flux (\ref{eq:F_ave}) and apparent luminosity.
The apparent luminosity is given by (\ref{eq:L_app}), while the actual one as $L$ is obtained by integration of the flux over the directions in the reference frame of a NS. 
The ratio of the apparent and actual luminosity is called the amplification factor and can be calculated as
\beq
\label{apparent_lum}
a=
\frac{L_{\rm app}}{L}=
\frac{4\pi F_{\rm ave}(i,\theta_B)}{\int_{4\pi} F_{\rm ave}(i,\theta_B){\rm d}\Omega}.
\eeq
Note, that we define an actual luminosity as averaged over distant observers. 
Such definition has an automatic correction for gravitational redshift. 
\item 
Investigation of statistical features.\\
We treat $i$ and $\theta_B$ as random values with a probability density 
\beq 
\rho(\theta)\propto\sin{\theta}.
\eeq  
It leads to the consideration of the apparent pulsar luminosity as a random value too. 
Calculating the apparent luminosity for various $i$ and $\theta_B$, we get the distributions of X-ray sources over the apparent luminosity.
The distributions are used to calculate the standard deviation and quantiles of 10 and 90 per cent of possible luminosity amplification $a$. 
We define the quantile of 90 per cent $L_{90}$ as the value below which the luminosities of 90 per cent of NS lie. 
By analogy, the quantile 10 per cent $L_{10}$ is the value below which lies 10\% of the apparent luminosity. 
Further, everywhere in the article, the actual luminosity is taken as one $L=1$.
\end{enumerate}
     

The consideration of sub-critical accretion is performed in the same manner. 
The polar cap of opening angle $\theta_{\rm pol}$ is split into equal rings over latitude on the NS surface. 
We suggest that luminosity is uniformly distributed over the cap.
The sequence of rays is emitted from each ring into certain angles $\alpha$ and $\varphi$. 
Here, $\alpha$ is the angle between the normal to the star surface and the ray propagation direction, $\varphi$ is the azimuthal angle measured in the tangent plane. 
Each ray carries away an energy equal to $I(\alpha)\cos{\alpha}\sin{\alpha}$, where $I$ is specific intensity. The final angle $\psi$, measured from the conventionally upper magnetic pole of the NS, is calculated taking into account the gravitational light bending. The beam energy corrected for the redshift is added to the energy received by the observer at this $\psi$.

\section{Results}
\label{sec:Results}

\begin{figure*}
	\centering 
    \includegraphics[width=18cm]{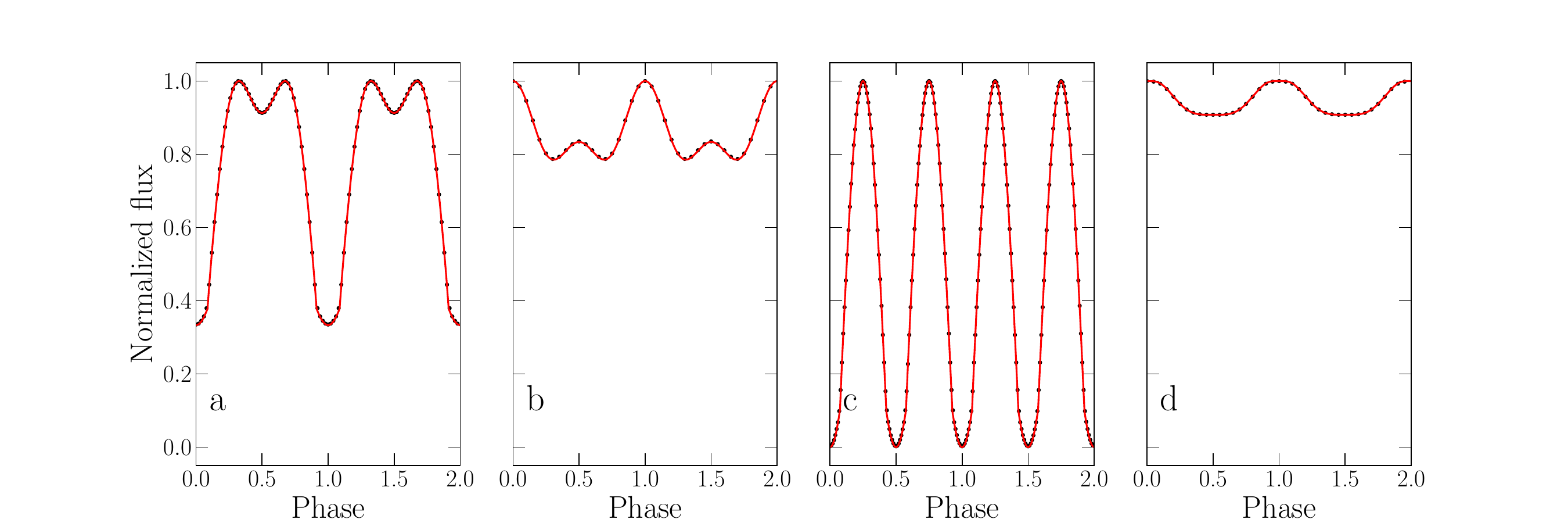} 	
	\caption{
       Theoretical pulse profiles in the case of a XRP with hot spots. 
       Solid lines correspond to our numerical simulation results, dashed ones correspond to the light curves specified by the analytical expression proposed by \citealt{2020A&A...640A..24P}. 
       Common parameters: $R_\mathrm{Sh}/R=0.5$, $\theta_\mathrm{pol}=0.1$~rad. Spesific parameters for each plot: 
       (a) b=-0.8, $i={\pi}/{3}$, $\theta_B={\pi}/{3.5}$; 
       (b) b=2.0, $i={\pi}/{5}$, $\theta_B={\pi}/{2.39}$; 
       (c) b=-1.0, $i={\pi}/{2}$, $\theta_B={\pi}/{2}$; 
       (d) b=0.5, $i={\pi}/{6}$, $\theta_B={\pi}/{3}$.
       All curves are normalized to the maximum of X-ray flux during the pulse period.
	}
	\label{pic:Pulse_profiles_spot}
\end{figure*}

\begin{figure*}
	\centering 	
 	\includegraphics[width=18cm]{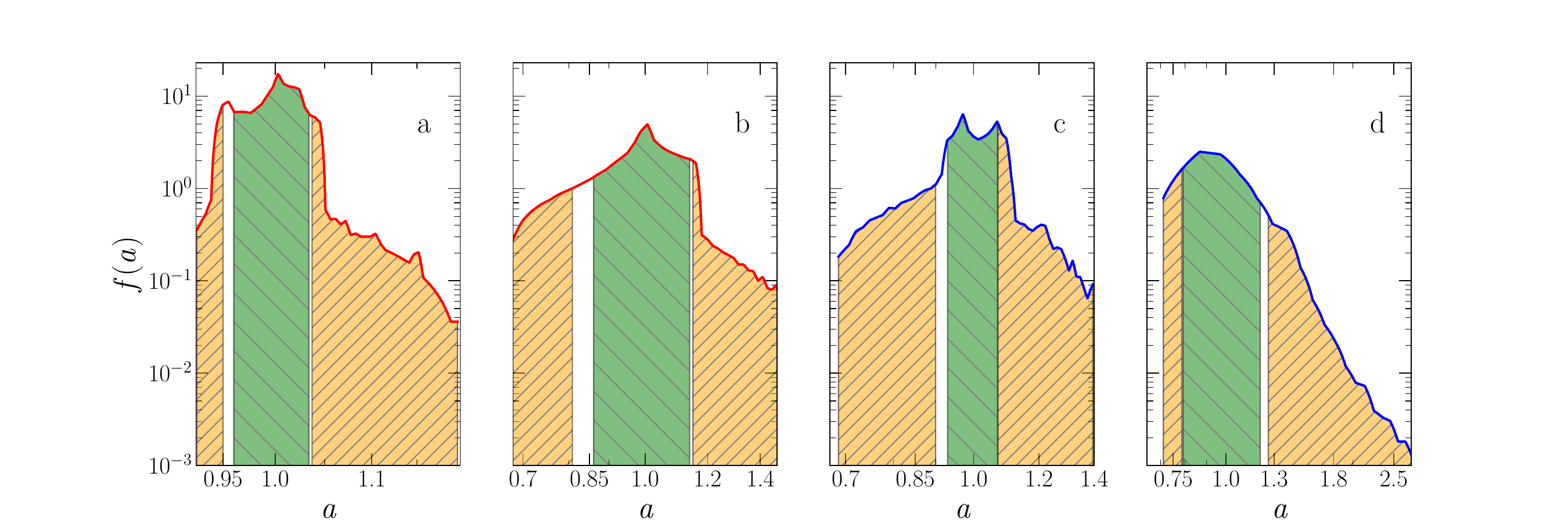} 	
	\caption{ 
   Distributions over the luminosity amplification
   factor $a$. \textit{All panels}: the orange area on the left shows a quantile of 10\%, the green one in the center shows $\overline{a}\pm\sigma(a)$, orange on the right -- a quantile of 90\%.  {Parameters:} $M=1.4M_\odot$, $R=12$~km. Panel a: a NS with a polar cap, $b=-0.2$, $\theta_{\rm pol}=0.1\,{\rm rad}$; b: a NS with a polar cap, $b=2.0$, $\theta_{\rm pol}=0.1\,{\rm rad}$; c: a NS with an accretion column, $H/R=0.25$, $r_c/R=0.05$, $g(h)={1}/{H}$, $\beta=\beta_\mathrm{ff}$; d: a NS with an accretion column, $H/R=11/12$, $r_c/R=0.05$, $g(h)={1}/{H}$, $\beta=\beta_\mathrm{ff}$.
    	}
	\label{pic:distrib}
\end{figure*}

\begin{figure*}
	\centering 
	\includegraphics[width=8cm]{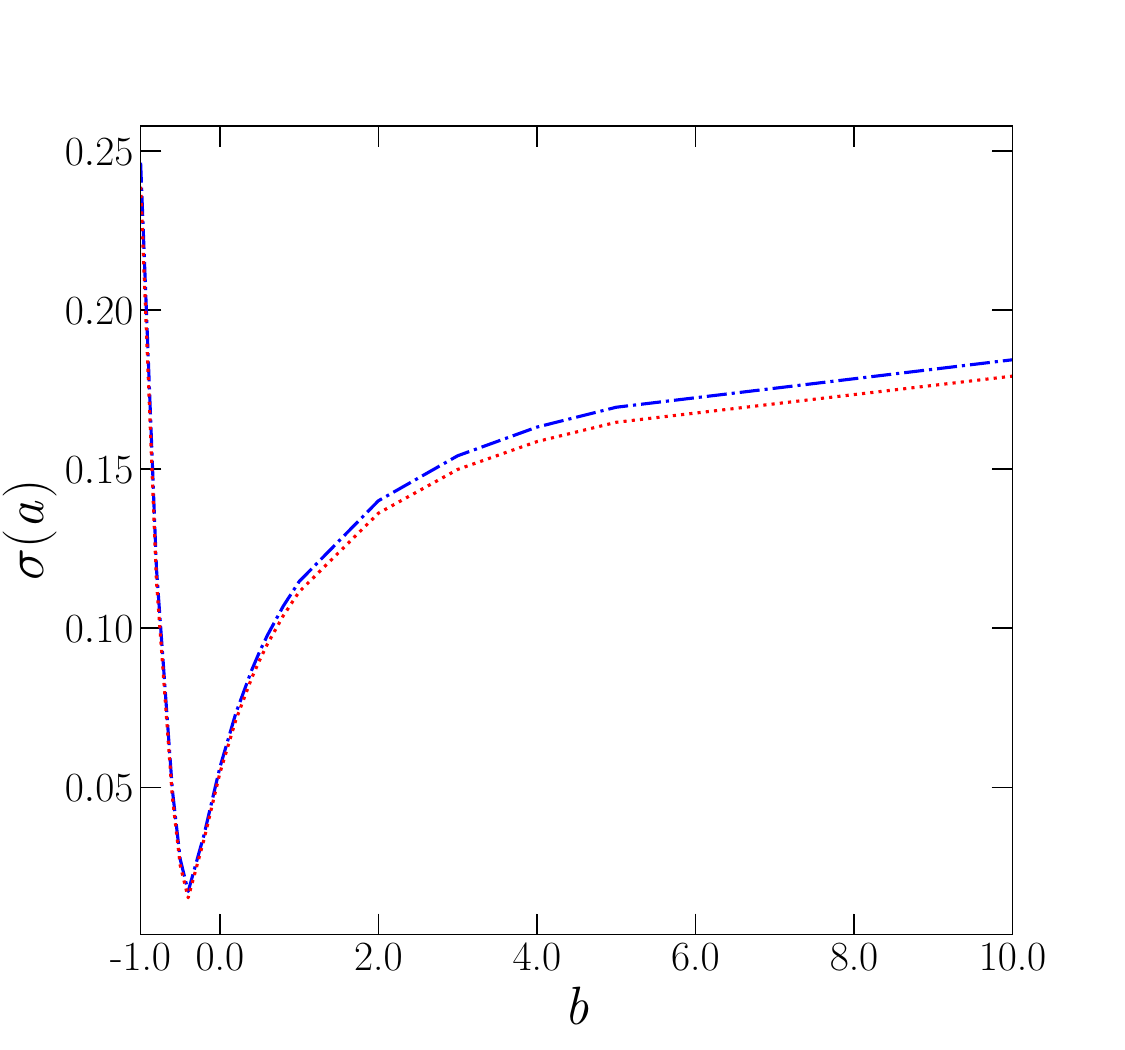} 	
 	\includegraphics[width=8cm]{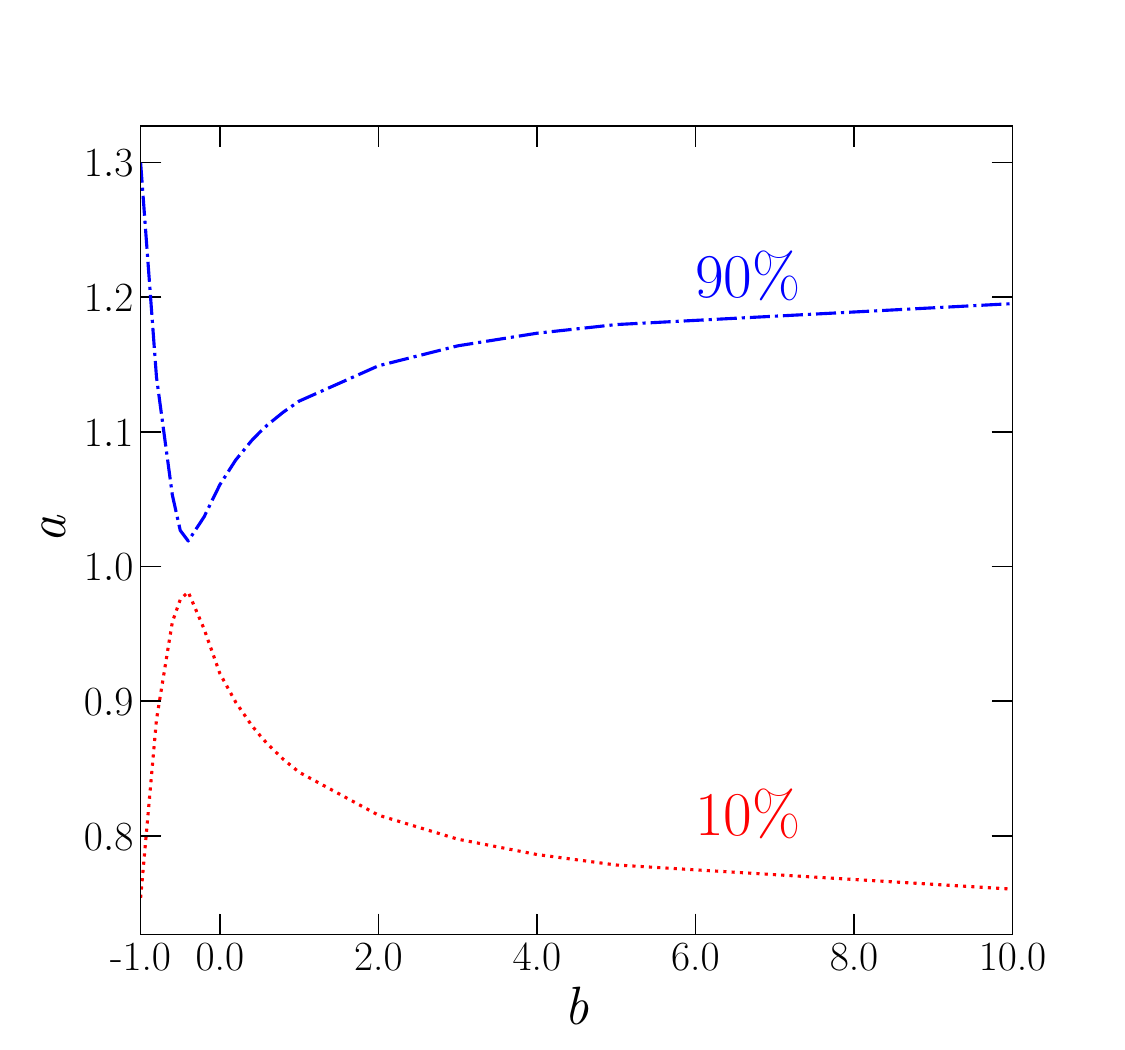} 	
	\caption{
{\it Left panel: }The dependence of amplification factor standard deviation on the beam pattern given by parameter $b$ (see equation \ref{eq:polar_cap_distr_fr}) for the case of sub-critical accretion.
Blue and red curves correspond to different size of hot spots: $\theta_{\rm pol}=0.05\,{\rm rad}$ and $0.2\,{\rm rad}$ respectively.
{\it Right panel: }Amplification factor quantiles 10\% (red dotted curve) and 90\% (blue dashed-dotted curve) as a function of beam pattern parameter $b$. {Parameters:} $M=1.4M_\odot$, $R=12$~km, $\theta_{\rm pol}=0.1\,{\rm rad}$.
	}
	\label{pic:L_sd_spot}
\end{figure*}

\begin{figure*}
	\centering 
	\includegraphics[width=8.cm]{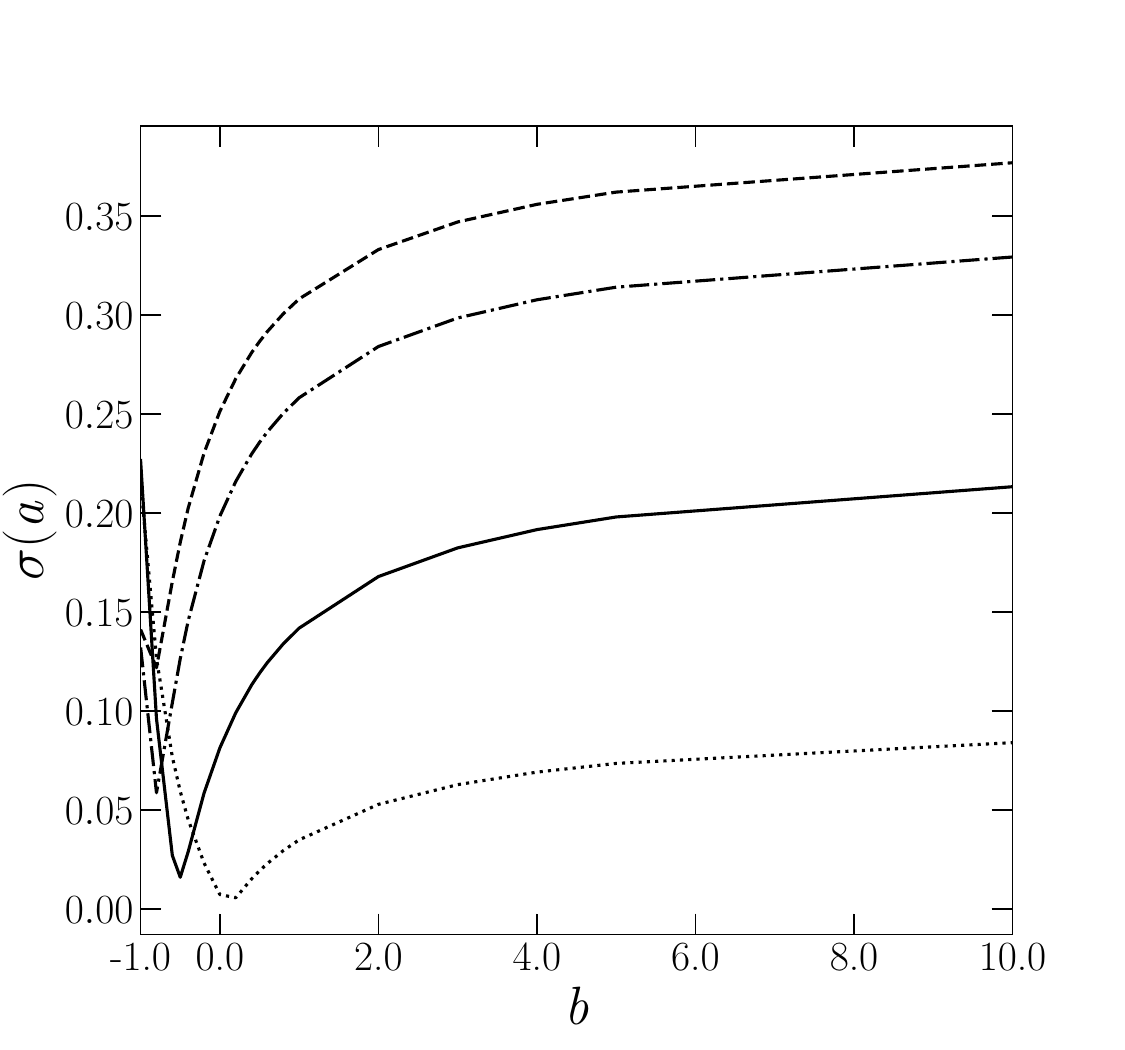} 		\includegraphics[width=8.cm]{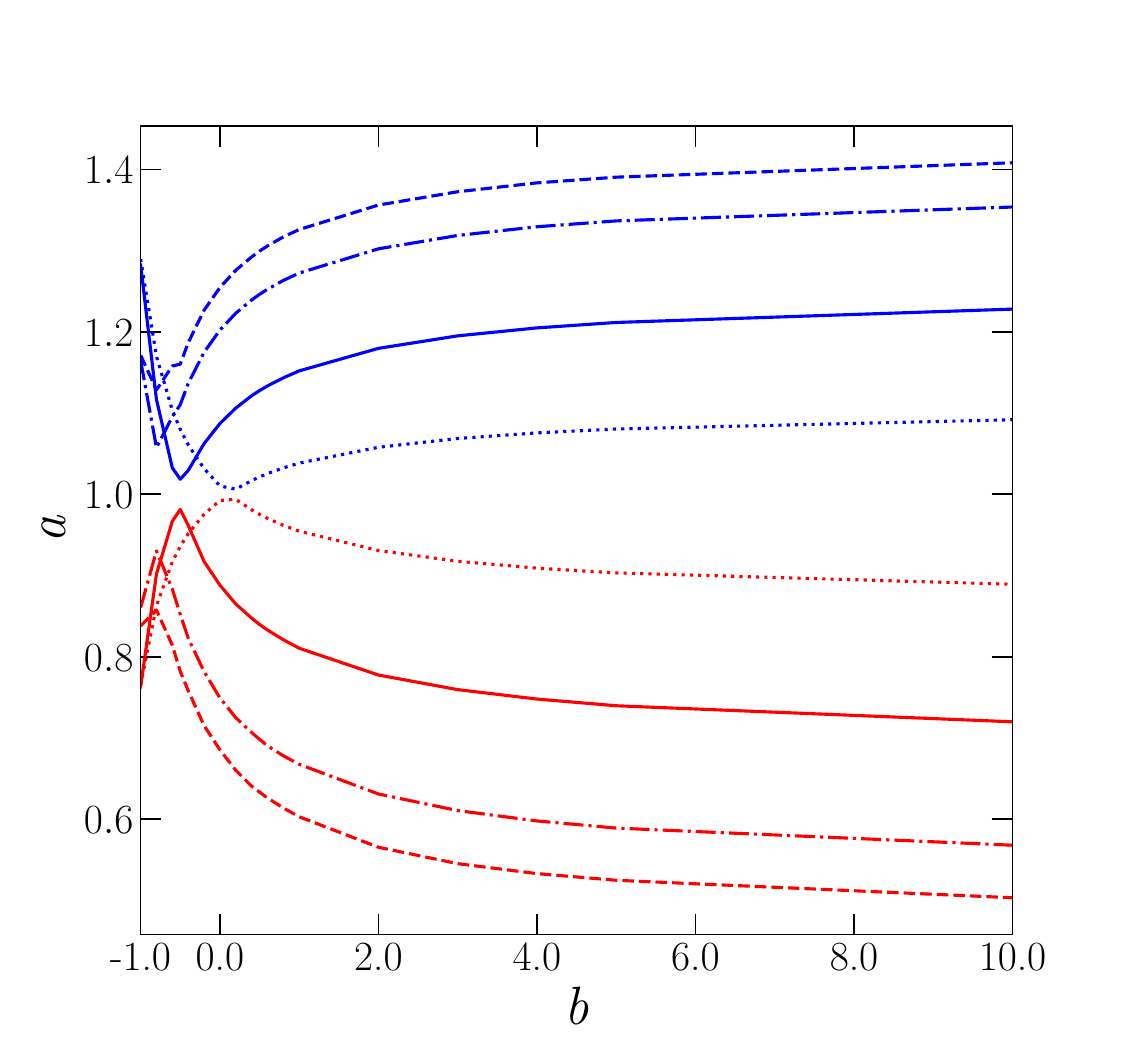}
	\caption{
 {\it Left panel: }
 The dependence of amplification factor standard deviation on the beam pattern given by parameter $b$ (see equation \ref{eq:polar_cap_distr_fr}) for the case of sub-critical accretion.
 Different curves correspond to different compactness of a NS: $R_\mathrm{Sh}/R=0.001$ (dashed), $R_\mathrm{Sh}/R=0.1$ (dash-dotted), $R_\mathrm{Sh}/R=0.3$ (solid), $R_\mathrm{Sh}/R=0.5$ (dotted).
{\it Right panel: } 
Amplification factor quantiles 10\% (red curves) and 90\% (blue curves) as a functions of beam pattern parameter $b$.
Different curves are given for different compactness of a NS: $R_\mathrm{Sh}/R=0.001$ (dashed), $R_\mathrm{Sh}/R=0.1$ (dash-dotted), $R_\mathrm{Sh}/R=0.3$ (solid), $R_\mathrm{Sh}/R=0.5$ (dotted).
	}
	\label{pic:L_10_spot_c}	
\end{figure*}

\begin{figure*}
	\centering 
	\includegraphics[width=8.cm]{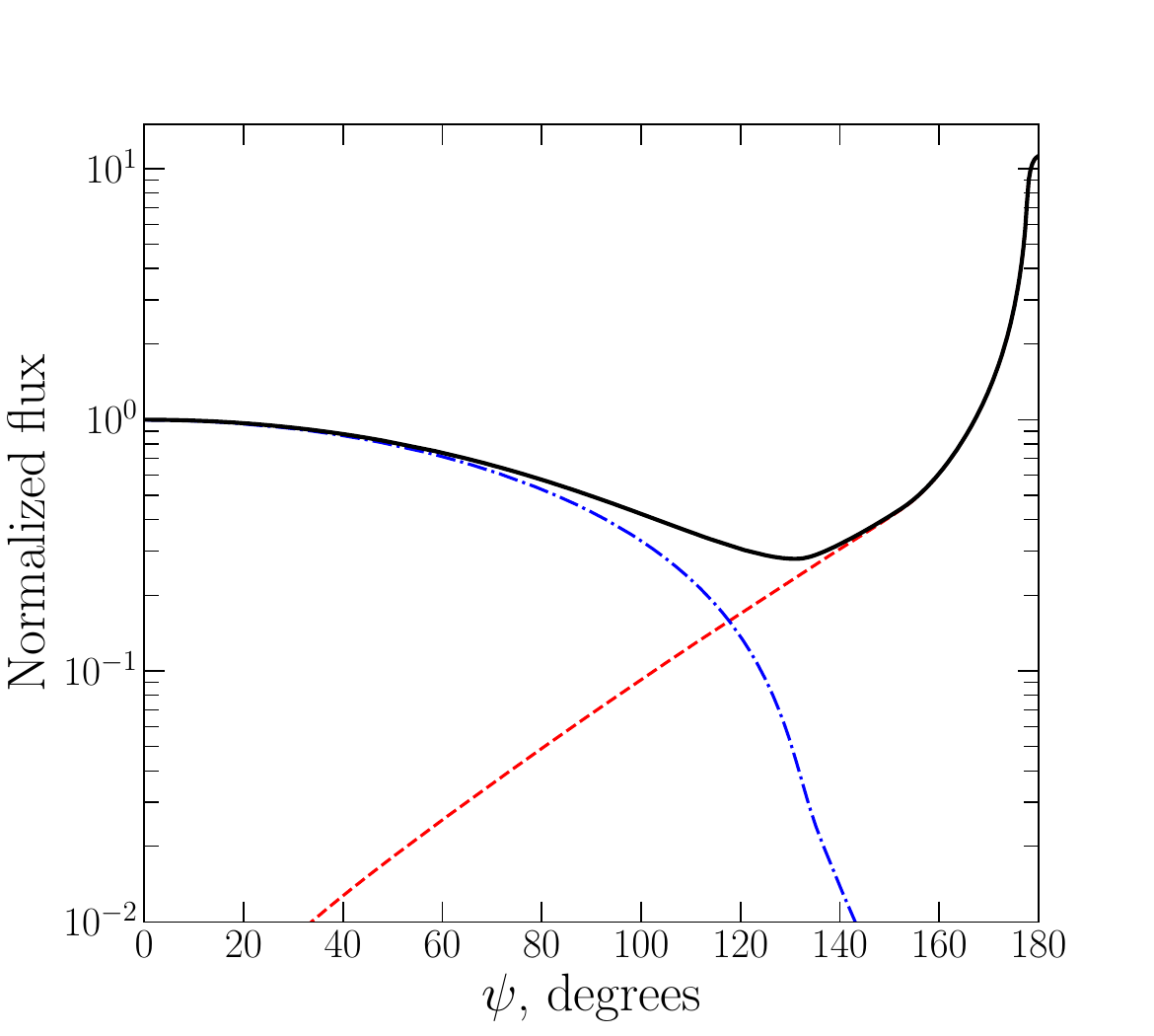} 
 	\includegraphics[width=8.cm]{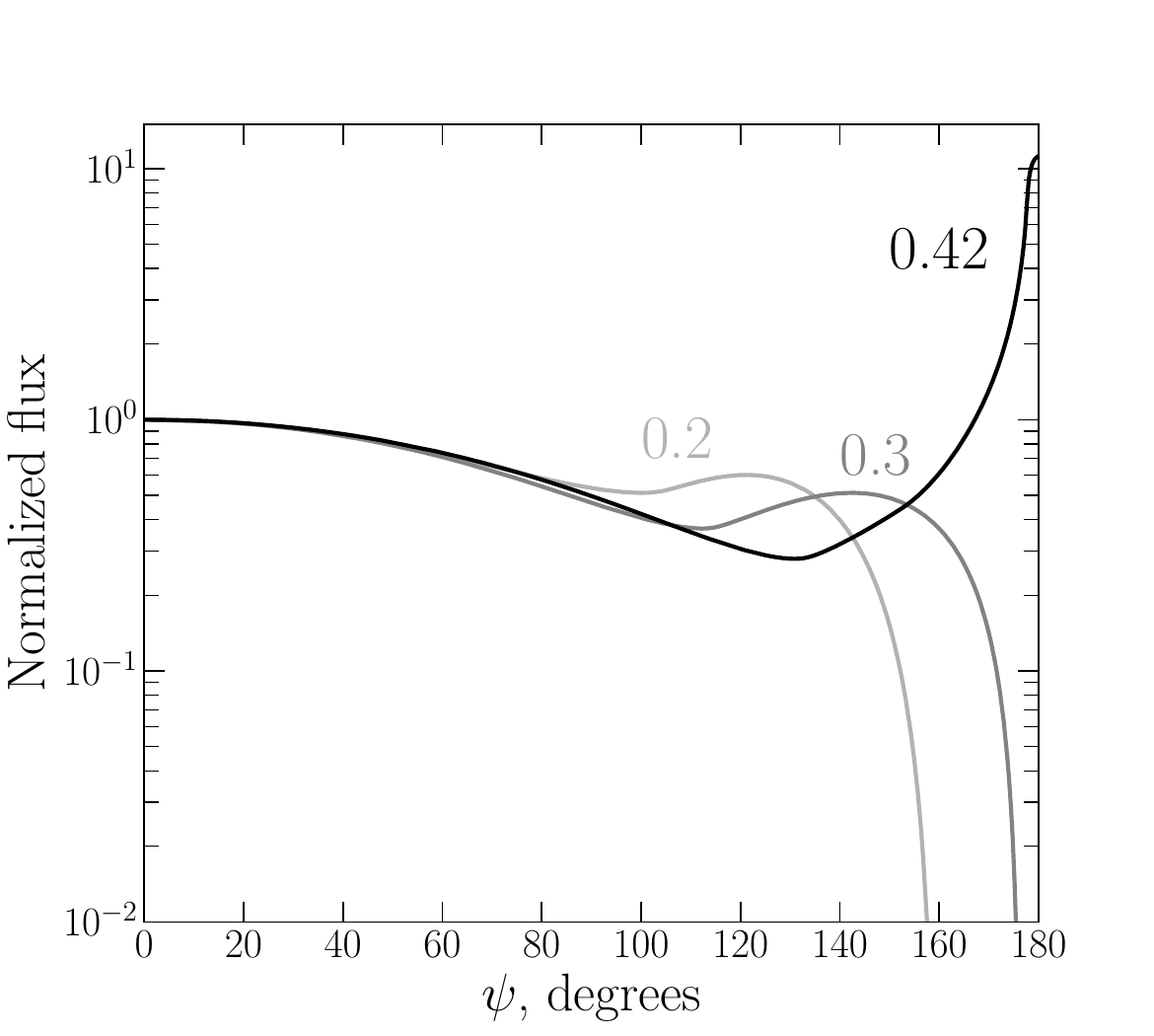} 	
	\caption{
        Angular distribution of the flux from a NS with one column. Parameters: $H/R=0.5$, $r_c/R=0.05$, $g(h)\propto(H-h)/(R+h)$, $\beta=\beta_\mathrm{ff}$
        {\it Left panel} shows the direct radiation from the column (red line), radiation reflected from the NS surface (blue line), and composition of two components (black line).
		NS compactness is fixed at $R_\mathrm{Sh}/R=0.42$.
  		{\it Right panel} shows angular distributions of the flux from a NS with one column calculated for different stellar compactness: 
        $R_\mathrm{Sh}/R=0.2$ ({light-grey} line), $0.3$ ({grey} line) and $0.42$ ({black} line).
	}
	\label{pic:Angle_flux1}
\end{figure*}

\begin{figure*}
	\centering 
    \includegraphics[width=18cm]{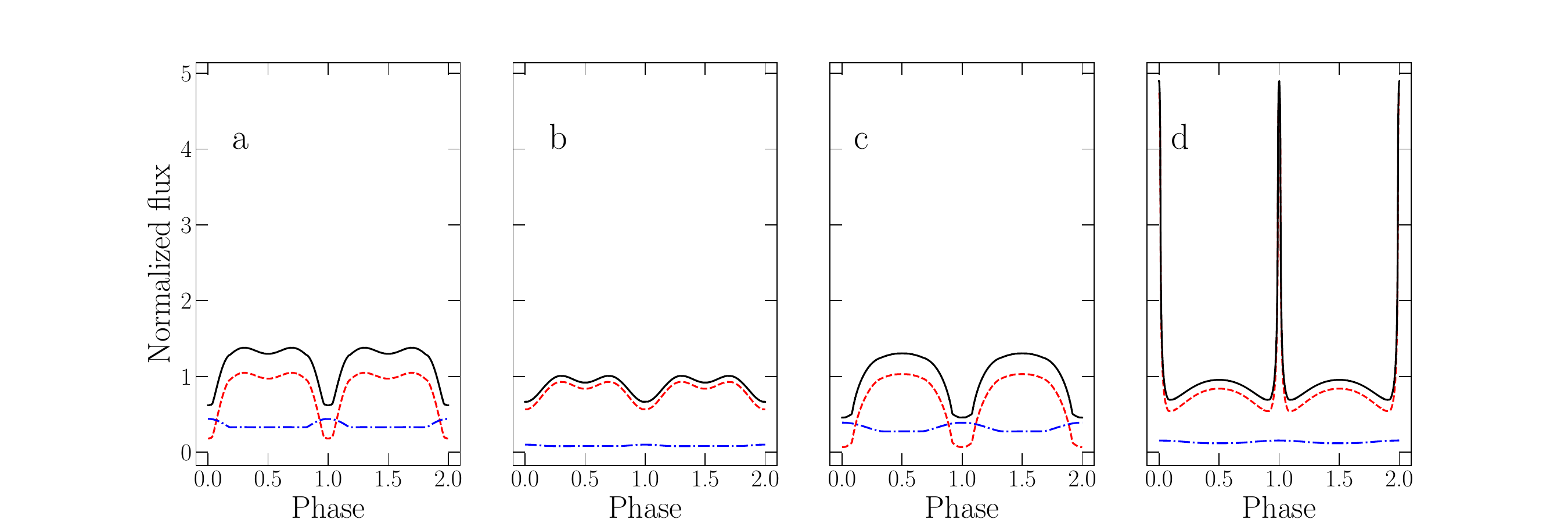} 	
	\caption{
       Theoretical pulse profiles in the case of a pulsar with accretion columns. Different lines are the decomposition of total radiation (black solid) into the component reflected from the surface (blue dashed-dotted line) and component visible directly (red dashed?) from accretion column.  
       Different plots correspond to different accretion column height, observer's inclination in respect to the XRP and magnetic obliquity: 
       (a) H/R=1/12, $i={\pi}/{2.5}$, $\theta_B={\pi}/{4}$; 
       (b) H/R=1.0, $i={\pi}/{2.5}$, $\theta_B={\pi}/{4}$; 
       (c) H/R=1/6, $i={\pi}/{5}$, $\theta_B={\pi}/{5}$; 
       (d) H/R=2/3, $i={\pi}/{6}$, $\theta_B={\pi}/{6}$ 
       The flux is normalized by the average value over all distant observers. 
       Parameters: $M=1.4\,M_\odot$, $R=12$~km, $\beta=0$, $r_c/R=0.05$, $g(h)=1/H$. \\
	}
	\label{pic:Pulse_profiles_column}
\end{figure*}

\begin{figure*}
	\centering 
	\includegraphics[width=8.cm]{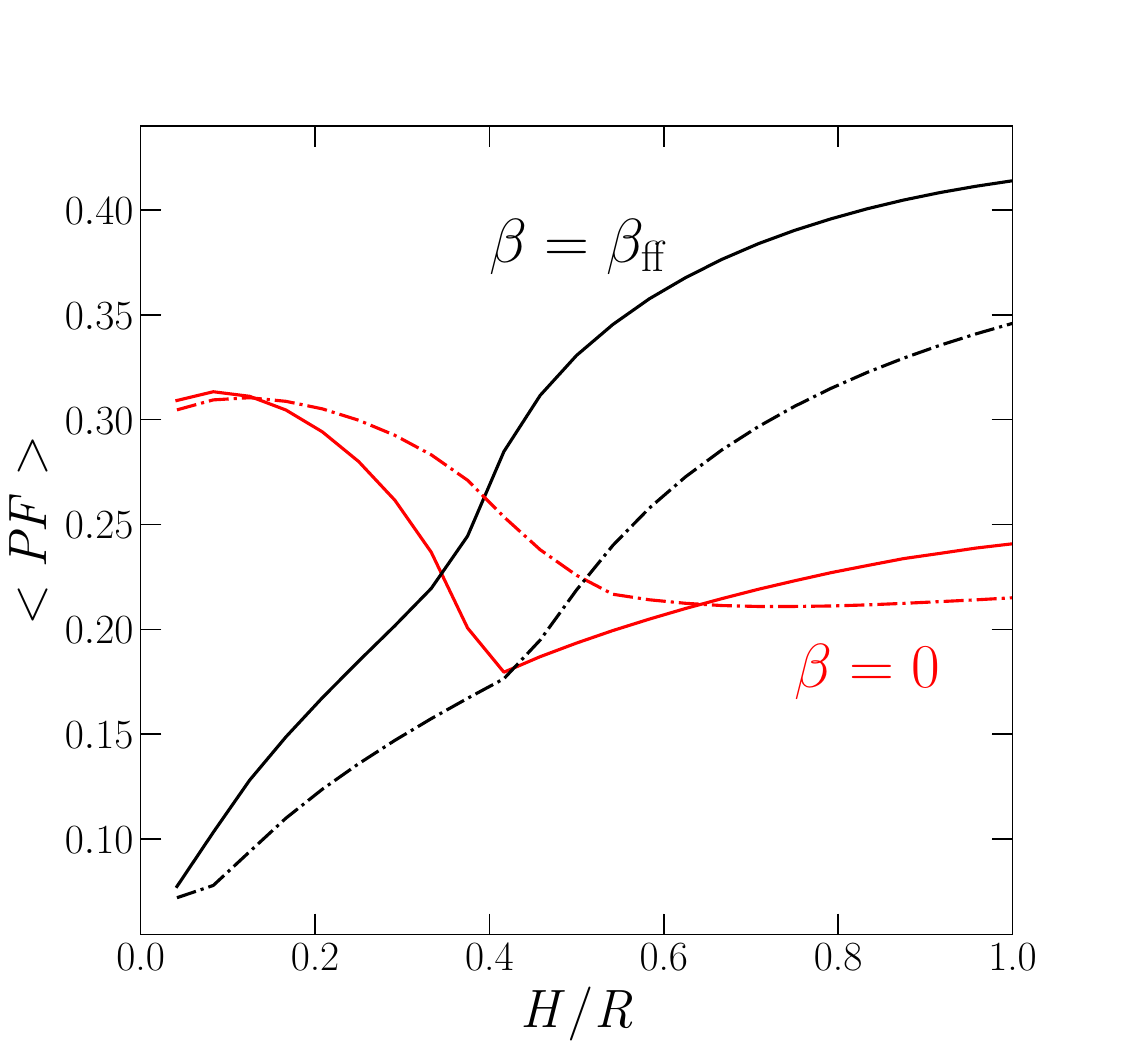} 	
 	\includegraphics[width=8.cm]{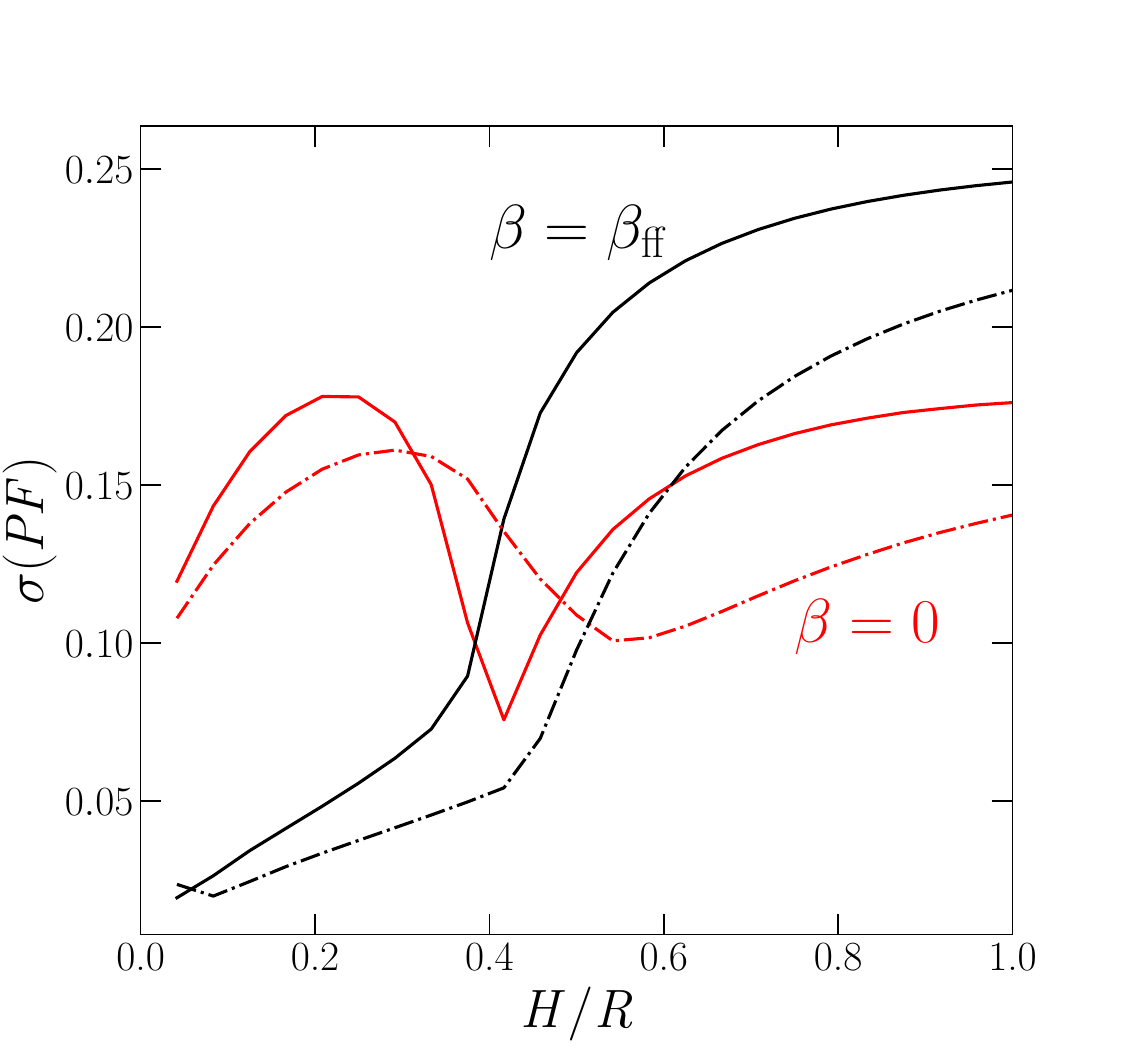} 	
	\caption{
    The dependence of the average pulsed fraction (left) and the standard deviation of a pulsed fraction (right) on the height of accretion column calculated for different column models. The solid black line: $\beta=\beta_{\mathrm{ff}}$, $g(h)={1}/{H}$; solid red line: $\beta=0$, $g(h)={1}/{H}$; dashed-dotted black line: $\beta=\beta_{\mathrm{ff}}$, $g(h)\propto{(H-h)}/{(R+h)}$; dashed-dotted red line: $\beta=0$, $g(h)\propto{(H-h)}/{(R+h)}$. 
    Common parameters: $M=1.4M_\odot$, $R=12$~km, $r_c/R=0.05$.
	}
	\label{pic:PF}
\end{figure*}

\subsection{Sub-critical accretion}

The pulse profiles for sub-critical pulsar, i.e. with a hot spot at the NS surface, can be obtained analytically using an approximate expression of light bending in the Schwarzschild metric \citep{2020A&A...640A..24P}.
They, therefore, can be used to verify our numerical results.

The comparison of the pulse profiles obtained by our numerical code and the pulse profiles obtained analytically is shown in Fig.\,\ref{pic:Pulse_profiles_spot}. 
All profiles in Fig.\,\ref{pic:Pulse_profiles_spot} are obtained for the case of NS compactness $R_\mathrm{Sh}/R=0.5$. 
It corresponds to the most massive known NS (see, e.g., \citealt{2021ApJ...918L..28M,2017A&A...608A..31N}) and allows us to test the accuracy of calculations under the conditions of strong gravitational bending. 
It is clearly seen that the numerical and analytical results agree well, which indicates that our model is sufficiently accurate.   

In the case of a hot spot, there are two main parameters of the simulations: the parameter describing the beam pattern $b$ (see eq.\,\ref{eq:polar_cap_distr_fr}) and a size of a polar cap given by the opening angle $\theta_{\rm pol}\sim 0.1\,{\rm rad}$. 
However, one can see (Figs. \ref{pic:L_sd_spot}) that the {standard} deviations of the apparent luminosity from the actual one are dependent on the polar cap size insignificantly.
Thus, we will treat only $\theta_{\rm pol}=0.1\,{\rm rad}$ in our further analysis.





Accounting for the gravitational bending effect significantly influences the distribution of sub-critical pulsars over the apparent luminosity (see Fig.\,\ref{pic:L_10_spot_c}).
The increasing of NS compactness leads to more isotropic distribution of apparent luminosity and to smaller errors in ones measuring.
Such behaviour is natural because, in the case of hot spot geometry, the gravitational bending results in the propagation of light in the larger solid angle than in the non-relativistic situation, which makes the beam pattern more isotropic. 
For example, when ${R_\mathrm{Sh}}/{R}=0.5$ and $b\approx0$, any possible mistakes in the luminosity estimation go to zero.

\subsection{Distribution functions}
\label{sec:Distr_func}
{Examples of distribution functions of the amplification factor $a$ are given in Figure \ref{pic:distrib}. The orange color on the left and right shows the areas limited by the quantile values of 10\% and 90\%, respectively. The green area demonstrates the deviation $\pm\sigma(a)$ around the average value. It can be seen that the boundaries of the standard deviations are close to the values of the quantiles. The distribution functions can have a complex, asymmetric bimodal structure. Since such behaviour is rather unclear, we carried out the verification of our distribution function calculations. The description of the verification process is in Appendix \ref{App:Verification_distribution}.
}

\subsection{Super-critical accretion}

Super-critical accretion results in appearance of accretion column - extended source of X-ray photons above NS surface.
A detailed analysis in this section is carried out for a NS with the following parameters: 
$$M=1.4\,M_\odot,\quad R=12~\text{km},\quad 
r_c/R=0.05,$$ 
where $M$ is the mass of the NS, $R$ is its radius and $r_c$ is the radius of accretion column base.
Note that these mass and radius of the NS correspond to $R_\mathrm{Sh}/R\approx0.35$, where $R_\mathrm{Sh}$ is the Schwarzschild radius. 
{In all simulations the height of the column varies in the range from $0$ to $1$ NS radius. 
}

\subsubsection{Pulsed fraction}

At super-critical state, not only the brightness but also the geometry of the emitting region become dependent on the mass accretion rate. 
The beam pattern changes as a function of luminosity and therefore one can expect noticeable changes in the pulse profile.
Total flux is composed of the direct flux from accretion columns and the flux reflected from the NS surface (see Fig.\,\ref{pic:Angle_flux1}). 
The examples of the theoretical pulse profiles in the case of super-critical XRPs are presented in the Fig.\,\ref{pic:Pulse_profiles_column}. 
Profiles are symmetric because the shape of the column base is assumed as full-filled circle or ring.
Note that the pulsar with $H/R=2/3$, $i=\theta_B=\pi/6$ demonstrates a huge amplification of flux when the line of sight is directed close to the magnetic axis. 
Such amplification requires not only the coincidence of the magnetic axis and the line of sight but also sufficient height of accretion column. 
The pulse profiles obtained for the super-critical case can be understand more clearly if we consider Fig.\,\ref{pic:Angle_flux1} which are demonstrate the dependence of the flux on the angle $\psi$ measured from the magnetic axis in the case of NS with only one column. 
Under the condition of sufficiently compact NS, there is a strong flux amplification on the opposite side of the NS relative to the column due to the gravitational lenzing of X-ray photons. 
The main contribution to the radiation at these angles is provided by direct flux from the column. 
In the opposite direction, the radiation flux is dominated by the component reflected from the stellar surface. 

Possibly strong anisotropy of X-ray energy flux in the case of super-critical XRPs results in a large pulsed fraction:
\beq\label{eq:PF}
{\rm PF} = \frac{F_{\rm max}-F_{\rm min}}{F_{\rm max}+F_{\rm min}},
\eeq 
where $F_{\rm min}$ and $F_{\rm max}$ are the minimal and the maximal X-ray energy flux over the pulse period.
The pulsed fraction depends on the geometry of emission region, i.e. on accretion column height, and NS rotation with respect to the distant observer. 
Averaging the pulsed fraction over possible mutual orientation of the observer and a NS, we conclude that the pulsed fraction tends to increase with the increase of accretion column height and, thus, the accretion luminosity (see left panel in Fig.\,\ref{pic:PF}).
Depending on the angular distribution of X-ray photons at the walls of accretion column, the typical pulsed fraction can reach values of $0.3$ (in the case of photons leaving the walls of accretion column predominantly in the direction orthogonal to the field lines) or above $0.4$ (in the case of column radiation beamed towards the NS surface due to the scatterings in accretion flow of high velocity, \citealt{1988SvAL...14..390L,2013ApJ...777..115P}). 
Typical range of the pulsed fraction values tends to increase with accretion column height as well (see right panel in Fig.\,\ref{pic:PF}).
The expected increase of a typical pulsed fraction with the luminosity of super-critical XRPs is in agreement with observational results reported in a few bright XRPs (\citealt{2010MNRAS.401.1628T,2022ApJ...938..149H,2022MNRAS.517.3354L,2023arXiv230414881S}).

\subsubsection{Standard deviation of the apparent luminosity}

\begin{figure}
	\centering 
	\includegraphics[width=8cm]{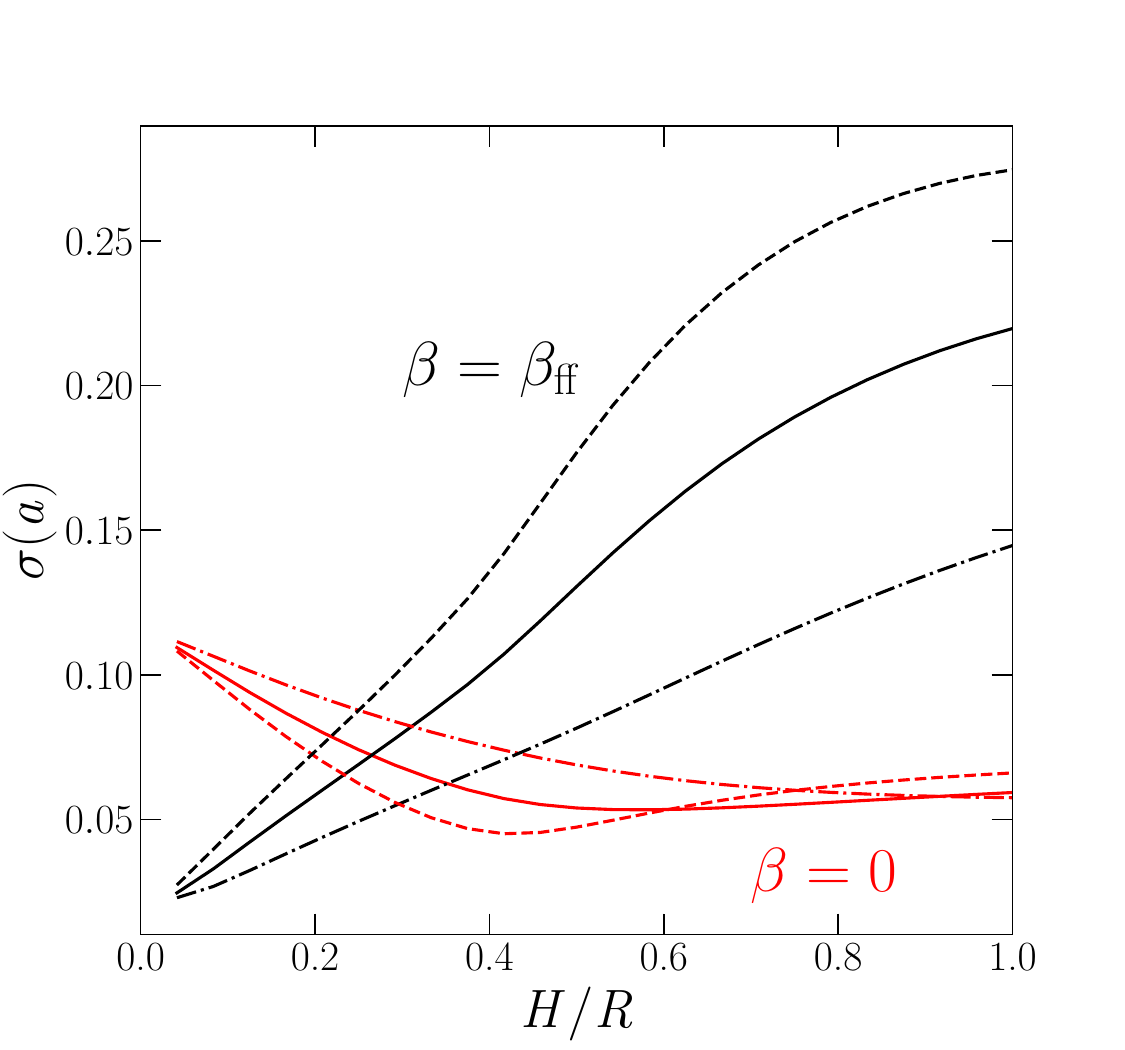} 	
	\caption{
		The standard deviation of the amplification factor. The solid line corresponds to a column of uniform luminosity $g(h)={1}/{H}$, dashed-dotted -- to a fractional linear one $g(h)\propto{(H-h)}/{(R+h)}$, dashed -- to $g(h)\propto h$. Red lines show $\beta=0$, black ones - $\beta=\beta_{\rm ff}$. 
		Common parameters: $M=1.4\,M_\odot$, $R=12$ km, $r_c/R=0.05$.
	}
	\label{pic:L_sd_comparison} 
\end{figure}

The typical difference between the actual and apparent luminosity in super-critical XRPs depends on the relative height of the accretion column, distribution of its brightness over the height and angular distribution of photons leaving accretion column walls (see Fig.\,\ref{pic:L_sd_comparison}). 
The standard deviation of the amplification factor tends to be relatively small ($\lesssim 0.1$) for the scenario when X-ray photons leave accretion column walls predominantly along the local normal to the walls (see red lines in Fig.\,\ref{pic:L_sd_comparison}).
If the flux from accretion column walls is directed towards the NS surface (due to the photon scattering by free-falling material, see, e.g. \citealt{1988SvAL...14..390L}), the standard deviation of the amplification factor increases with accretion column height and can be as high as $\sim 0.3$. 

\begin{figure*}
	\centering 
	\includegraphics[width=8.cm]{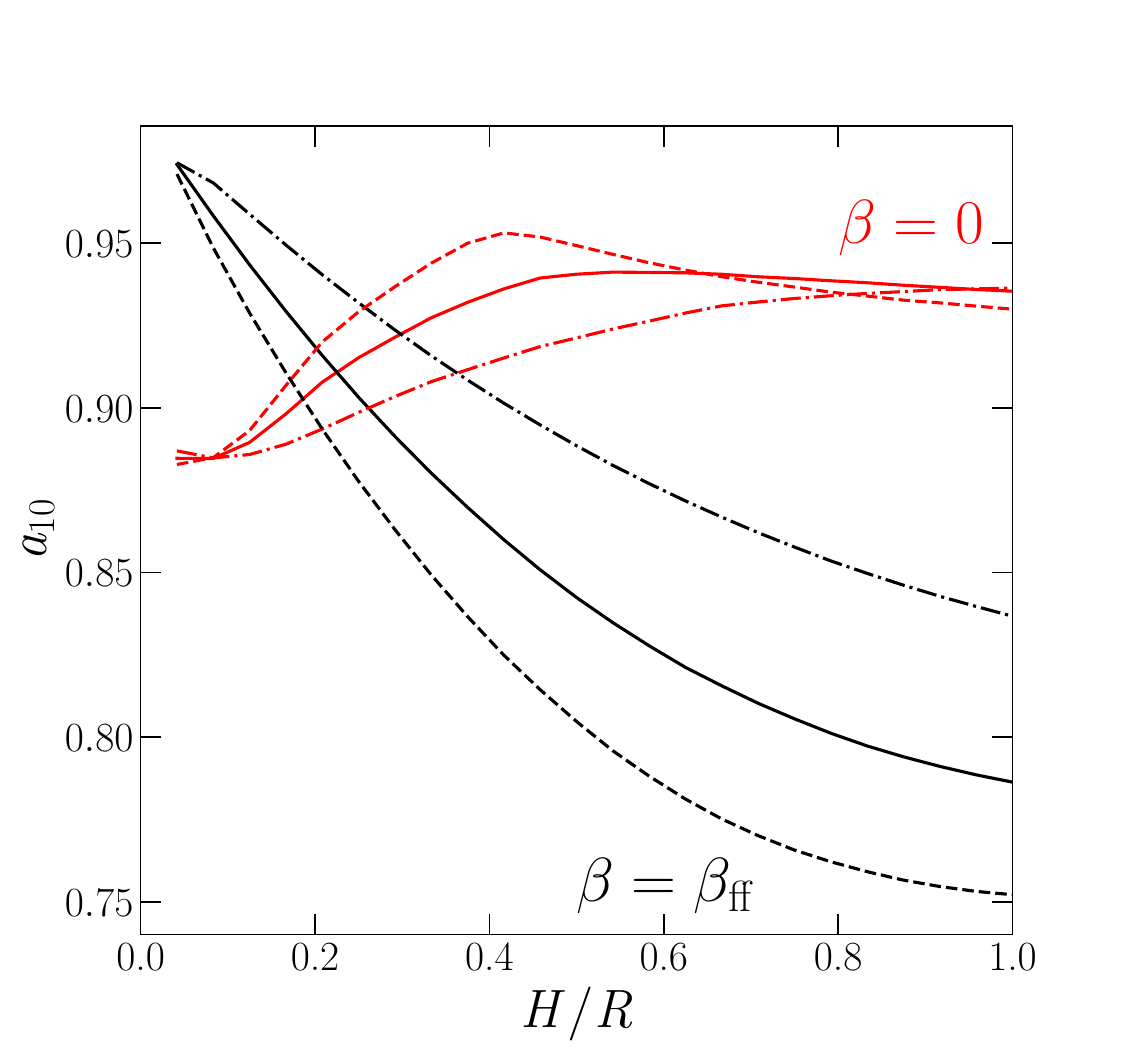} 	
 	\includegraphics[width=8.cm]{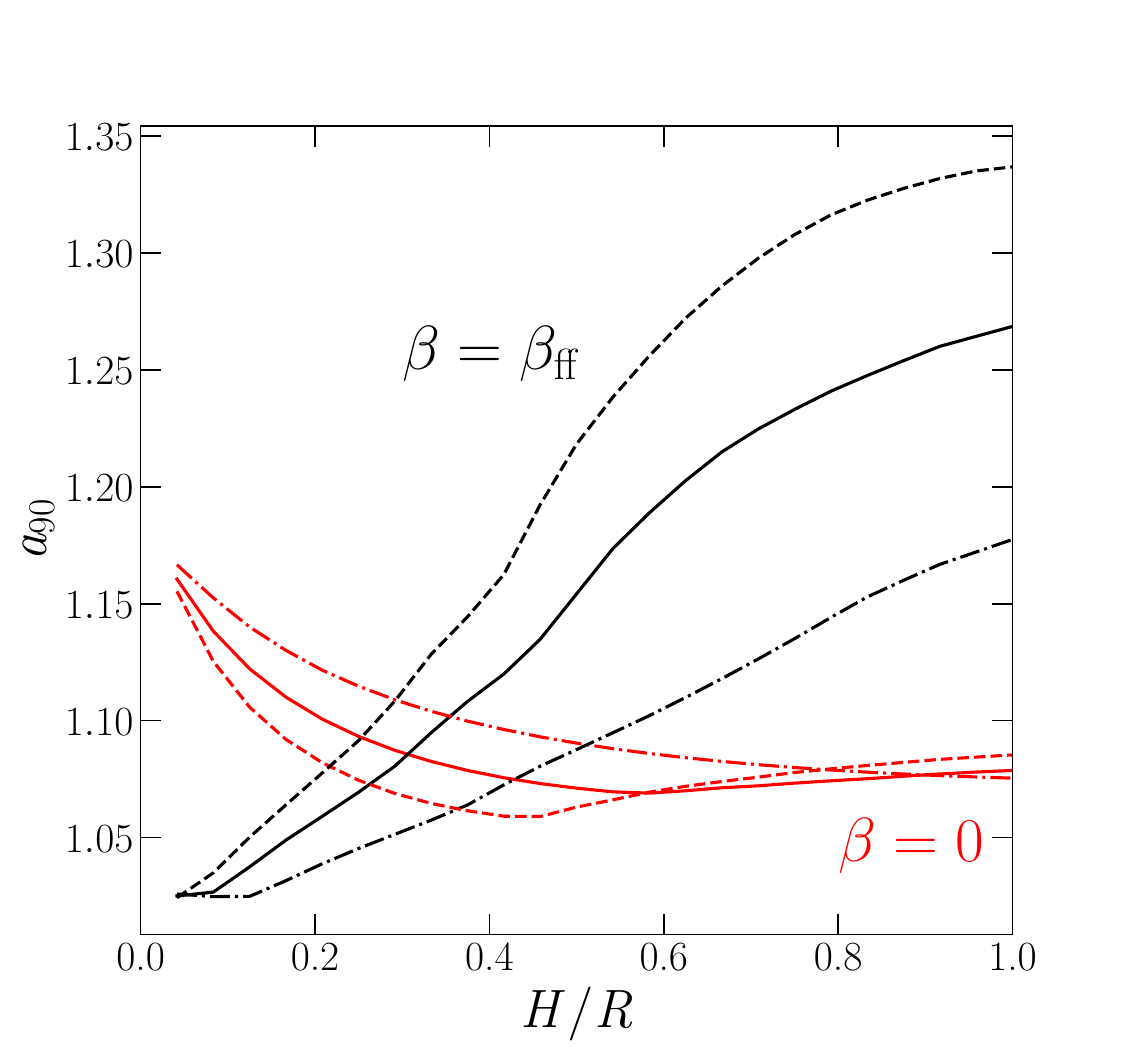} 	
	\caption{
        Amplification factor quantile 10\% (left panel) and 90\% (right panel) calcualted for super-critical XRP.
		The solid line corresponds to a column of uniform luminosity $g(h)={1}/{H}$, dashed-dotted -- to a fractional linear one $g(h)={1}/{H}\propto{(H-h)}/{(R+h)}$, dashed -- to $g(h)\propto h$. Red lines show $\beta=0$, black ones - $\beta=\beta_{\rm ff}$. 
		Common parameters: $M=1.4\,M_\odot$, $R=12$ km, $r_c/R=0.05$.
	}
	\label{pic:L_10_90}	
\end{figure*}

The behaviour of a standard deviation is consistent with quantiles $10\%$ and $90\%$ (see Fig.\,\ref{pic:L_10_90}).
It is also seen that the apparent luminosity of 90\% of sources lies above $L_\mathrm{app} \approx0.8L$ and below $L_\mathrm{app}\approx1.3L$.

\subsubsection{Influence of NS compactness}

\begin{figure}
	\centering 
	\includegraphics[width=8.5cm]{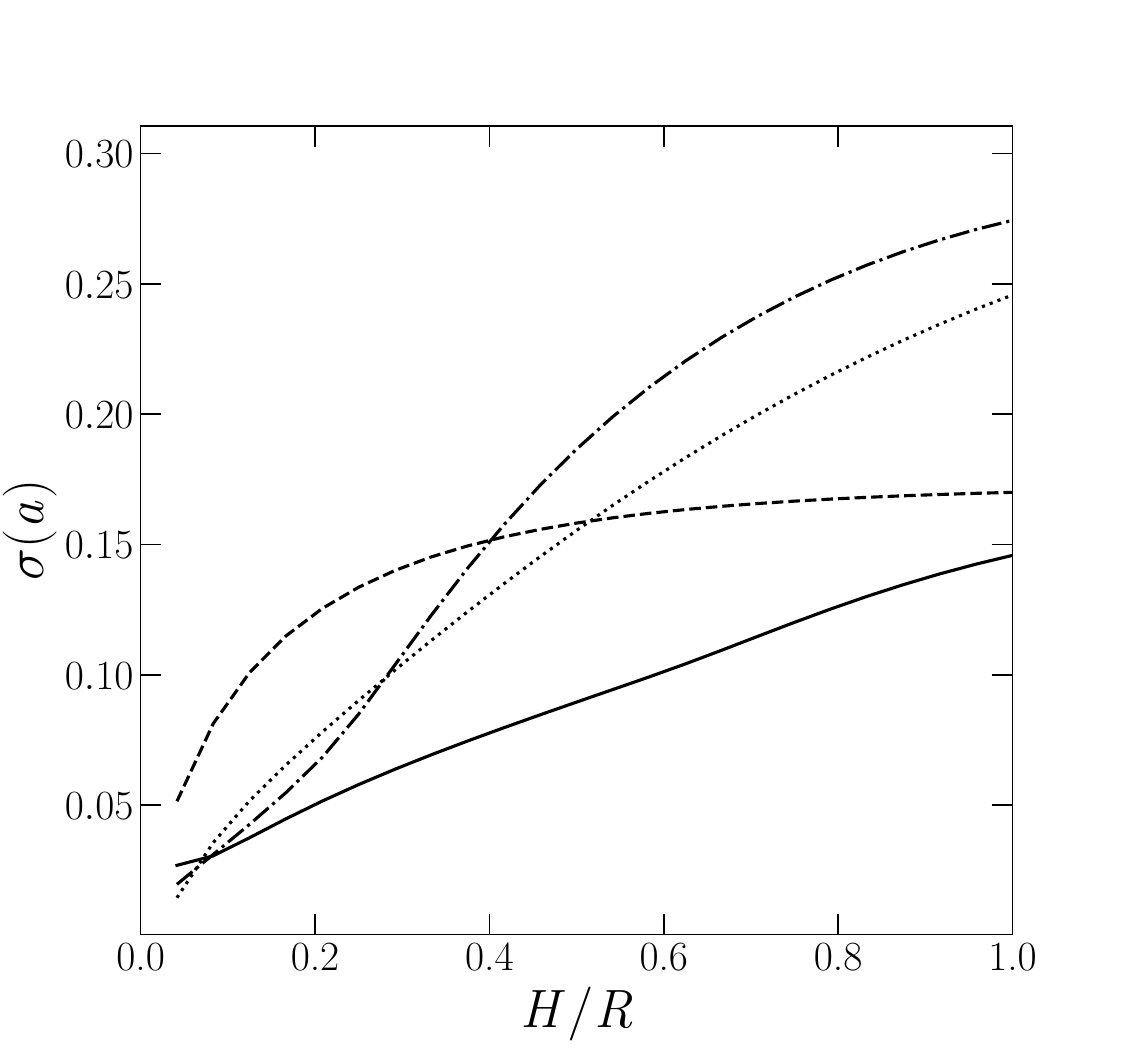} 	
	\caption{
 {The dependence of apparent luminosity standard deviation for the case of super-critical accretion with following parameters: $g(h)\propto{1}/{H}$, $\beta=\beta_\mathrm{ff}$, $r_c/R=0.05$.
 Different curves correspond to different compactness of a NS: $R_\mathrm{Sh}/R=0.001$ (dashed), $R_\mathrm{Sh}/R=0.3$ (solid), $R_\mathrm{Sh}/R=0.4$ (dashed-dotted), $R_\mathrm{Sh}/R=0.5$ (dotted).
        }
	}
	\label{pic:AC_compactness}
\end{figure}

The compactness of a NS influences the amplification factor for different distant observers.
The standard deviation of the amplification factor tends to be larger for NSs of larger $R_{\rm Sh}/R$ ratios (see Fig.\,\ref{pic:AC_compactness}).
While the standard deviation for the non-relativistic case ($R_{\rm Sh}/R\ll 1$) does not exceed $\sim 0.15$, for $R_{\rm Sh}/R$ it can be $\sim 0.3$ already.

\section{Discussion and Summary}

{
Estimation of accretion luminosities in XRPs based on averaging the X-ray energy flux over the pulse period may differ from the actual luminosity. 
This is especially a case for bright XRPs with accretion columns, where we can expect strongly anisotropic beam patterns of ray radiation.
We have investigated possible deviations of the apparent luminosity from the actual one assuming a simplified model of accretion column and accounting for possible reflection of X-ray flux from the NS surface and effects of gravitational bending in the vicinity of a NS.
To estimate typical difference between actual and apparent luminosity we have simulated the pulse profiles for different mutual orientation of NS rotation and a distant observer.
The apparent luminosities obtained from the pulse profiles were averaged over the random distant observers and parameters of NS rotation (see Section\,\ref{sec:alg}). 
}

In the case of polar cap geometry, the {standard} deviation of the apparent luminosity from the actual one is strongly dependent on the beam pattern forming at the NS surface (see Fig.\,\ref{pic:L_sd_spot}): the standard deviation of the apparent luminosity is relatively small (i.e., $\sigma(L_{\rm app})/L\sim 0.05$) at beaming pattern parameter $b\approx 0$, and increases towards larger $b$ up to $\sigma(L_{\rm app})/L\sim 0.2$.

In the case of super-critical XRPs with accretion columns, {a typical difference between the apparent and actual luminosities increases with a rise of accretion column height} (see Fig.\,\ref{pic:L_sd_comparison}), i.e. the typical difference is expected to be larger at higher accretion luminosity.
For the accretion columns that produce radiation beamed towards NS surface, the {relative standard} deviation can be as high as 20\% (see black lines in Fig.\,\ref{pic:L_sd_comparison}).
In 90\% of cases the amplification factor $a\in (0.8,1.25)$ for the case of accretion columns 
(see Fig.\,\ref{pic:L_10_90}), i.e.
\beq
0.8L\lesssim L_{\rm app}\lesssim 1.25L.
\eeq 

Such behaviour is explained by the fact that a fraction of accretion column radiation becomes a subject of a strong gravitational lensing by a NS (see Fig.\,\ref{pic:Angle_flux1}).
Note, that the beaming of accretion column radiation at its walls influence significantly the typical difference between the actual and apparent luminosity (compare black and red lines in Figs.\,\ref{pic:L_sd_comparison} and \ref{pic:Angle_flux1}).
The {standard} deviation of the apparent luminosity is affected by the compactness of a NS (see Fig.\,\ref{pic:AC_compactness}). 
The increase of compactness results in the rise of possible deviations for super-critical pulsar. 
On the other hand, for NSs with a hot spot the difference between the apparent and actual luminosity tend to decrease with the growth of compactness.

{
Investigating the pulsed fraction dependence on the accretion column height, we conclude that it tends to increase with the increase of accretion column height and, thus, the accretion luminosity.
For the case of high accretion columns, the pulsed fraction can be above $0.4$ (see left panel in Fig.\,\ref{pic:PF}). 
The increase of a typical pulsed fraction with the luminosity of super-critical XRPs is in agreement with observational results reported in a few bright X-ray transients: 
V~0332+53 (see Fig.\,10 in \citealt{2010MNRAS.401.1628T}),
RX~J0209.6-7427 \citep{2022ApJ...938..149H},
SMC~X-3 \citep{2022MNRAS.517.3354L},
and RX~J0440.9+4431 (see Fig.\,5 in \citealt{2023arXiv230414881S}).
In the case of RX~J0440.9+4431, increase of the pulsed fraction is observed up to luminosity $L\sim 6\times 10^{38}\,\ergs$, while at higher luminosities the pulsed fraction decreases (see Fig.\,7 in \citealt{2022ApJ...938..149H}). 
The pulsed fraction reduction at $L>6\times 10^{38}\,\ergs$ can be due both to the accretion column becoming sufficiently high (see Fig.\,\ref{pic:PF}) and to the fact that at high mass accretion rates, the disc tends to lose matter due to the radiation driven winds, which results in geometrical collimation of X-rays and decrease of the pulsed fraction (see, e.g., \citealt{2023MNRAS.518.5457M}).
}

\section*{Acknowledgements}

This work was supported by the Russian Science Foundation grant 19-12-00133-P (IDM) and
UKRI Stephen Hawking fellowship (AAM).
We are grateful to Alexander Salganik, Sergey Tsygankov and an anonymous referee for their useful comments which helped improve the paper.

\section*{Data availability}

The calculations presented in this paper were performed using a private code developed and owned by the corresponding author. All the data appearing in the figures are available upon request.


{

}

\appendix

\section{Verification of distribution function calculations}
\label{App:Verification_distribution}

\begin{figure}
	\centering 
	\includegraphics[width=8.5cm]{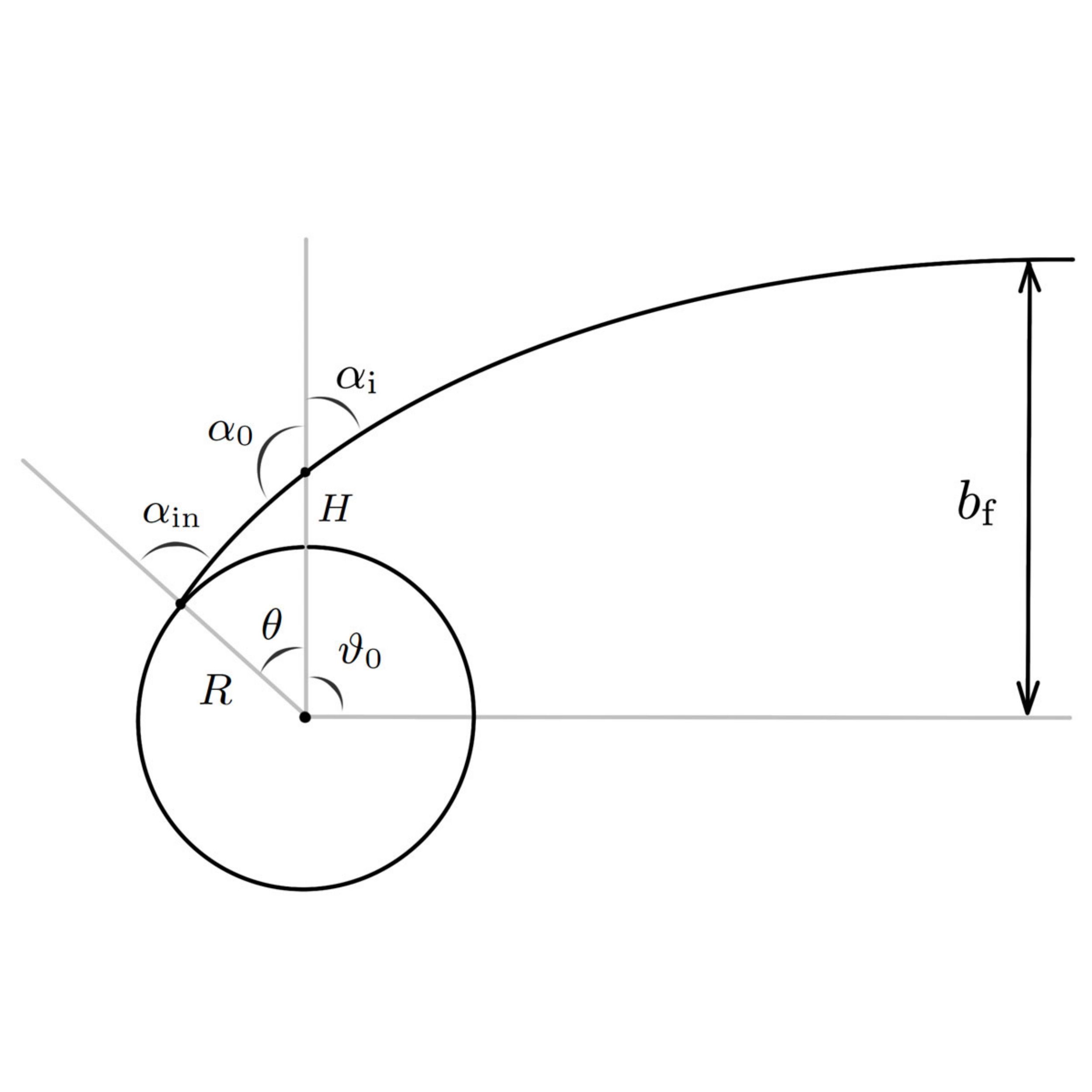} 	
	\caption{
Geometry of light bending in the Schwarzschild metric. The black solid curve shows the photon trajectory connecting the stellar surface at radius $R$ and co-latitude $\theta$ and the point at distance $r = R + H$ located above NS surface. 
	}
	\label{pic:scheme_2}
\end{figure}

Distribution functions calculated in the section \ref{sec:Distr_func} may have a complicated bimodal structure. 
Such behaviour is rather non-intuitive, so we perform the verification of our calculations of the luminosity distribution function. 
To do this, we compare the results of our program with the distribution function obtained for the model of a NS with two opposite hot spots neglecting general relativity effects and assuming that spots are flat. 
It corresponds to the results of our model with the compactness parameter $R_\mathrm{Sh}/R=0.001$ and spot opening angle $\theta_\mathrm{pol}=0.1$~rad (see \ref{sec:Results}). 

{The intensity of one spot is given by:
\beq
\label{eq:polar_cap_intensity}
	 I(\alpha)=I_0(1+b\cos\alpha),
\eeq
where $\alpha$~-- is the angle between the outer normal to the stellar surface and radiation direction. The normalization condition: 
\beq 
\label{eq:I0}
I_0=\frac{L}{4\pi(0.5+{b}/{3}) \Delta S},
\eeq where $\Delta S$ is the area of one spot, $L$ is the actual luminosity of a NS. 
The flux detected by the observer at a distance of $D$ is $F(\alpha)=I(\alpha){\Delta S}/{D^2}$. The flux averaged over all observers at this distance is 
$\overline{F}=(4\pi)^{-1}{L}/{D^2}$. 
Therefore, the ratio of the apparent NS luminosity and the actual one is 
\beq
\label{eq:a1}
a(\alpha)=\frac{F(\alpha)}{\overline{F}}=\frac{(1+b\cos{\alpha})\cos{\alpha}}{0.5+{b}/{3}}.
\eeq
}

{Let us consider a NS with fixed orientation angles $i$ and $\theta_B$. In this case
\beq
\label{eq:cosinus_theorem}
\cos{\alpha}=\cos i\cos\theta_B +\sin i\sin\theta_B\cos\varphi_{\rm p},
\eeq here $\varphi_{\rm p}$~-- is the phase angle. After averaging over the pulse profile the value of $a$ is given by 
\beq
\label{eq:a2}
a&=&{\frac{1}{2\pi L}\int\limits_0^{2\pi} L_\mathrm{app}(i,\theta_B,\varphi_{\rm p})\d\varphi_{\rm p}}\nonumber \\
&=&{\frac{1}{2\pi}\int\limits_0^{2\pi}\frac{F(\alpha(i,\theta_B,\varphi_{\rm p}))}{\overline{F}}\d\varphi_{\rm p}}.
\eeq
Because of the symmetry, we can limit the range of possible angles to $i,\theta_B\in[0,{\pi}/{2}]$. 
There are two possible cases. 
In the first one, the observer sees both spots alternately during the rotation of a NS. It is occured when $\cot{\theta_B}\cot{i}\leq1$. On the other hand, when $\cot{\theta_B}\cot{i}>1$, the observer always sees only the nearest hot spot. In the first case the value of the factor $a$ is 
\beq
\label{eq:a3}
a=\frac{C_2}{\pi}(\pi-2\varphi_0)\cos{\theta_B}\cos{i}+\frac{2C_2}{\pi}\sin{\theta_B}\sin{i}\sin{\varphi_0}\\ +bC_2\cos^2{\theta_B}\cos^2{i}+\frac{bC_2}{2}\sin^2{\theta_B}\sin^2{i},\nonumber
\eeq
while in the second one 
\beq
\label{eq:a4}
a=C_2\cos{\theta_B}\cos{i}+bC_2\cos^2{\theta_B}\cos^2{i}\\
 +\frac{bC_2}{2}\sin^2{\theta_B}\sin^2{i}.\nonumber
\eeq
Here $C_2=(0.5+{b}/{3})^{-1}$ and $\varphi_0=\arccos({\cot{\theta_B}\cot{i}})$.
}

{Having an analytical formula for $a$ at certain values $\theta_B$ and $i$ a distribution function can be obtained by the follow way. It have to set the random sample of remote observers isotropically located in space, which means that angles $\theta_B$ and $i$ have distribution function $\rho(\theta_B,i)\propto\sin({\theta_B,i})$. For each of these observers the factor $a$ is calculated through analytical formulas. after that the histogram of $a$ normalized to the unit area of the subgraph is created.
}

\begin{figure}
	\centering 
    \includegraphics[scale=0.43]{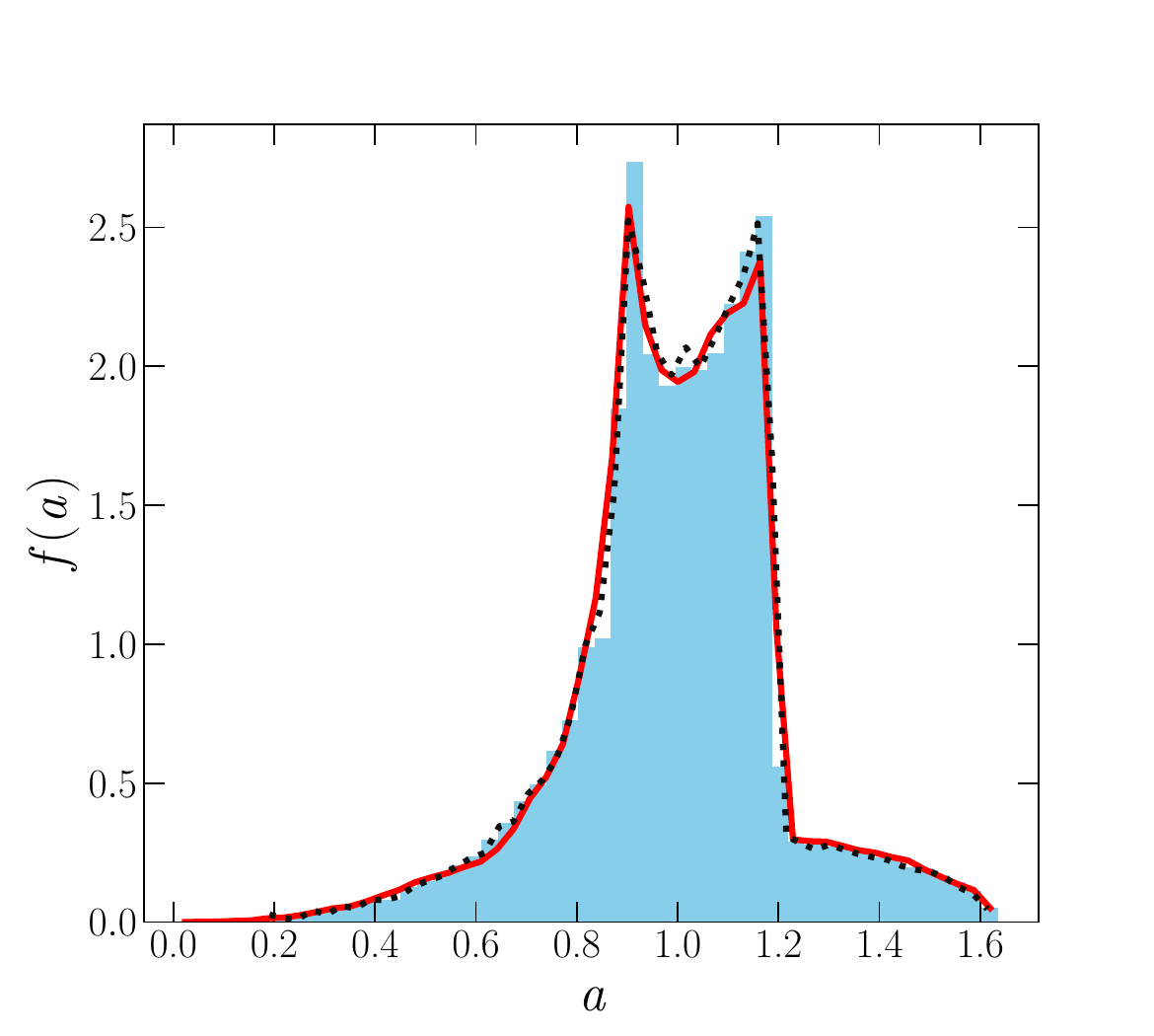} 	
	\caption{
      An example of a distribution function in the case of a star with a polar cap and non-relativistic compactness, calculated in three different ways. The black dotted curve corresponds to the calculations performed on the basis of our model ($R_\mathrm{Sh}/R=0.001$), the blue area is the calculation of the distribution function through analytical formulas, the red curve is an intermediate case in which averaging over the pulse profile was performed numerically, and the flux was calculated analytically. Parameter $b=-0.4$.
	}
	\label{pic:Distr_check}
\end{figure}

{The Fig.\,\ref{pic:Distr_check} demonstrates the example of the procedure described above. The blue area is the histogram obtained with the usage analytical formulas. The black dotted line corresponds to the calculations performed by our main program with the parameters of the NS $R_\mathrm{Sh}/R=10^{-3}$ and $\theta_\mathrm{pol}=0.1$~rad. The red line is an intermediate case, when averaging along the pulse profile is performed numerically, but at the same time the model of the spot is considered in neglecting of its curvature and the influence of general relativity effects, which allows us to immediately write an expression for the flux received by the observer at a specific time. } 

It is seen that all three curves are in a good agreement that confirms the correctness of the distribution function calculations of the parameter $a$ based on our model results. Note that there is a bimodal structure on the present graph.

\section{Accounting for the light bending in the Schwarzschild geometry}
\label{App:GR}

In the case of the Schwarzschild geometry, the photon trajectories belong to the planes formed by the radius-vector connecting the point mass to the trajectory point and the velocity vector at this point.
Let us use polar coordinates $(r',\varphi')$ to describe the trajectories. 
Then trajectory of a photon is described by the differential equation \citep{1973grav.book.....M}:
\beq 
\frac{\d^2 u}{\d\varphi'^2} = \frac{3}{2    }\,u^2 - u,
\eeq
where $u\equiv R_{\rm Sh}/r'$.

{
In this paper, we used our own code calculating approximate trajectories of X-ray photons in in close vicinity of a NS.
To approximate the photon trajectory, we use the following approximation formula  \citep{2020A&A...640A..24P}:
\begin{equation}\label{eq:light_bend}
x\simeq\left(1-u_{\rm i}\right) y\left(1+\frac{u_{\rm i}^2y^2}{112}-\frac{e u_{\rm i}y}{100}\left[\ln\left(1-\frac{y}{2}\right)+\frac{y}{2}\right]\right),
\end{equation}
where $x=1-\cos\alpha_{\rm i}$, $y=1-\cos\vartheta_0$, $\alpha_{\rm i}$ is the angle between the radius vector of the photon emission point and photon initial momentum, $\vartheta_0$ is the lensing angle (see Figure \ref{pic:scheme_2}), $u_{\rm i}$ is the ratio of the Schwarzschild radius to the length of the radius vector of the photon emission point. 
We call the lensing angle the angle between the direction of the photon motion at an infinite distance from the star and the radius vector of the photon trajectory initial point.} 

{
The formula \ref{eq:light_bend} allows us to extract the initial angle $\alpha_{\rm i}$ from known the lensing angle $\vartheta_0$ and the height $H$ of the photon above the surface of a NS. In our calculations, we are faced with the opposite problem: we have to obtain the lensing angle $\vartheta_0$ as a function of known $\alpha_{\rm i}$ and $H$. To do this we solve the transcendental equation \ref{eq:light_bend}. It is carried out by Newton's iterative method. We use the approximate value of $\vartheta_0$ obtained by the formula from \citep{2002ApJ...566L..85B} as an initial guess for Newton's method. 
}

{
If the ray goes to infinity without crossing the surface of a NS, then, knowing the lensing angle $\vartheta_0$ and the position of the ray starting point, we calculate the angle between the magnetic axis of the NS and the direction of propagation of the light beam at infinity based on trivial geometric formulas. The situation is more complicated for rays that crossing the NS surface. In this case we have to find the latitude of the photon landing point. Let us consider the photon that is emitted at a height of $H$ at an angle of $\alpha_0$ to the accretion column and intersects the surface of the NS at co-latitude $\theta$ at an angle of $\alpha_\mathrm{in}$ to the normal to the surface of the star (see \ref{pic:scheme_2}). Let's add the photon trajectory to the right for the starting point. A ray of light emitted at an angle of $\pi-\alpha_0=\alpha_{\rm i}$ to the accretion column would go along such a way. The lensing angle $\vartheta_0$ for the added trajectory can be found as the solution of equation \ref{eq:light_bend}.  For this trajectory, the parameter $b_{\rm f}$ is equal to \citep{2020A&A...640A..24P}:
\beq
b_{\rm f}=\frac{R+H}{\sqrt{1-{R_\mathrm{Sh}}/{(R+H)}}}\sin{\alpha_{\rm i}}
\eeq
If we know $b_{\rm f}$, we can express $\alpha_\mathrm{in}$, because they are related by the next formula:
\beq
b_{\rm f}=\frac{R}{\sqrt{1-{R_\mathrm{Sh}}/{R}}}\sin{\alpha_\mathrm{in}}
\eeq
The figure shows that for a beam emitted at an angle of $\alpha_\mathrm{in}$, the lensing angle is $\vartheta=\vartheta_0+\theta$. We can find the angle $\vartheta$ the equation \ref{eq:light_bend}, because we know $\alpha_\mathrm{in}$. Therefore, we can obtain latitude $\theta=\vartheta-\vartheta_0$ without numerical integration of the geodesic line equation in the Schwarzschild metric. 
}

\bsp	
\label{lastpage}
\end{document}